\documentclass[12pt,preprint]{aastex}

\usepackage{subfigure}
\usepackage{color}

\usepackage{graphicx}
\usepackage{amsmath,amssymb,amsfonts}
\usepackage{times}
\usepackage{here}
\def\vector#1{\mbox{\boldmath $#1$}}  

\shorttitle{}
\shortauthors{Kawabata et al.}

\begin{document}


\title{Extrapolation of Three-dimensional Magnetic Field Structure in Flare-Productive Active Regions with Different Initial Conditions}


\author{Y. Kawabata\altaffilmark{1}, S. Inoue \altaffilmark{2}, and T. Shimizu \altaffilmark{3, 4}}
\affil{National Astronomical Observatory of Japan}

\email{kawabata.yusuke@nao.ac.jp}


\altaffiltext{1}{National Astronomical Observatory of Japan, 2-21-1 Osawa, Mitaka, Tokyo 181-8588, Japan}
\altaffiltext{2}{Institute for Space-Earth Environmental Research, Nagoya University, Furo-Cho, Chikusa-ku Nagoya 464-8601, Japan}
\altaffiltext{3}{Institute of Space and Astronautical Science, Japan Aerospace Exploration Agency, 3-1-1 Yoshinodai, Chuo, Sagamihara, Kanagawa 252-5210, Japan}
\altaffiltext{4}{Department of Earth and Planetary Science, The University of Tokyo, 7-3-1 Hongo, Bunkyo-ku, Tokyo 113-0033, Japan}
\begin{abstract}
The nonlinear force-free field (NLFFF) modeling has been extensively used as a tool to infer three-dimensional (3D) magnetic field structure.
In this study, the dependency of the NLFFF calculation with respect to the initial guess of the 3D magnetic field is investigated. 
While major part of the previous studies used potential field as the initial guess in the NLFFF modeling, we adopt the linear force-free fields with different constant force-free alpha as the initial guesses.
This method enables us to investigate how unique the magnetic field obtained by the NLFFF extrapolation with respect to the initial guess is.
The dependence of the initial condition of the NLFFF extrapolation is smaller in the strong magnetic field region. Therefore, the magnetic field at the lower height ($< 10$ Mm) tends to be less affected by the initial condition (correlation coefficient $C>0.9$ with different initial condition), although the Lorentz force is concentrated at the lower height. 
\end{abstract}


\keywords{Sun: flares --- Sun: corona --- Sun: magnetic fields }


%
\section{Introduction}
Solar active regions contain the strong magnetic field and sometimes produce explosive events, such as solar flares and coronal mass ejections (CMEs) by releasing the magnetic energy.
The magnetic reconnection releases the non-potential magnetic energy, heats and accelerates the plasmas.
What is the condition on the onset of the solar flares and CMEs is one of the important open questions for solar flares and CMEs.
 Inferring correct 3D magnetic field is a crucial task to answer the question.

The magnetic field in the solar atmosphere is usually measured through the spectropolarimetric observations with slit-based or filter-based instruments.
The state-of-art spacecrafts, such as {\it Hinode}  \citep{2007SoPh..243....3K}  and {\it Solar Dynamics Observatory} \citep[{\it SDO};][]{2012SoPh..275....3P} have enabled us to observe magnetic field in the photosphere with high spatial resolution and high polarimetric accuracy. 
While photospheric magnetic field can be inferred with high accuracy through spectropolarimetric observations, it is difficult to obtain magnetic field in the corona.
The coronal lines have low brightness, are optically thin,  and suffer from larger Doppler broadening. 

The force-free field modeling is  one of the alternative methods to infer the 3D magnetic field in the solar corona.
The main concept of the force-free field modeling is to extrapolate the coronal magnetic field from the spatial map of the magnetic field in the photosphere  based on two assumptions \citep{2012LRSP....9....5W}. 
First assumption is the mechanical equilibrium of the plasmas in the solar corona.
Second assumption is the domination of the Lorentz force in the solar corona.
In the solar corona, the plasma  $\beta (=8\pi  p/B^2)$, which is the ratio between the plasma pressure and the magnetic pressure, is thought to be sufficiently small, $\beta \ll 1$ \citep{2001SoPh..203...71G}. 
The above two assumptions lead to the condition that the Lorentz force vanishes in the solar corona, 
\begin{equation}
\vector{j}\times\vector{B}=0,
\label{sec1:fff}
\end{equation}
where $\vector{j}$ is the current density.
The current density follows the Amp\'ere's law,
\begin{equation}
\nabla\times \vector{B}=\frac{4\pi}{c} \vector{j}.
\label{sec1:ampere}
\end{equation}
Equation (\ref{sec1:fff}) can be written
\begin{equation}
\nabla \times \vector{B}=\alpha(\vector{r})\vector{B},
\label{sec1:nlfff}
\end{equation}
where $\alpha$ is called the force-free parameter.
Generally, $\alpha$ is not a constant in space and such magnetic field distribution is called nonlinear force-free field (NLFFF).
 
 There exist mainly two approaches to solve NLFFF equations.
First approach is to solve the NLFFF equations with the boundary conditions, which is a mathematically well-posed problem.
The example of this approach was proposed by \cite{grad1958hydromagnetic}, and so is often called Grad-Rubin method.
In this approach, the distribution of $\alpha$ in only one polarity and the vertical magnetic field are prescribed to the bottom boundary.
One weak point of this approach is that the vector magnetic field on the bottom boundary is not consistent with the observed photospheric magnetic field.
Second approach is to find the closest force-free equilibrium field matching the observed vector magnetic field in the photosphere, which is prescribed to the bottom boundary.
While this approach keeps the bottom boundary consistent with the observed magnetic vector in the photosphere, this approach is an ill-posed problem and there is no proof of unique and stable solutions.
The examples of this approach are the  optimization methods \citep{2000ApJ...540.1150W, 2006A&A...457.1053W} and the magnetohydrodynamics (MHD) relaxation methods \citep{1981JCoPh..41...68C,1994ASPC...68..225M, 2014ApJ...780..101I}.

The NLFFF applications to the solar observations have some uncertainties, e.g., observational errors and numerical methods.
The question is,  in practical way,  how large the uncertainty of the NLFFF modeling is.
In other words, is there a possibility that completely different NLFFF results are obtained based on the same bottom boundary?
\cite{2006SoPh..235..161S} applied six NLFFF algorithm to analytical force-free field solutions and compared solutions with each other.
The best method reproduced the total energy in the magnetic field within 2\% error.
\cite{2008SoPh..247..269M} also compared NLFFF methods by using solar-like reference model. 
They applied NLFFF methods to both forced (not force-free) and force-free bottom boundaries.
They show that while the NLFFF with the force-free bottom boundary reproduced helical flux bundle, that with forced bottom boundary did not reproduce reference model well.
\cite{2009ApJ...696.1780D} applied various kinds of NLFFF algorithm (Grad-Rubin, optimization, and MHD relaxation) to the photospheric magnetic field observed with Hinode and investigated how different the solutions from different algorithm are.
They found that field lines from NLFFF  do not agree with EUV images and the average misalignment is 20-40 degrees.
\cite{2013ApJ...769...59T} investigated the effect of the observational instrument on the solution of NLFFF.
They used a photospheric magnetic field observed with different instrument, i.e. , the {\it Hinode}/SOT SP and {\it SDO}/Helioseismic and Magnetic Imager \cite[HMI;][]{2012SoPh..275..229S}.
They concluded that the relative estimates such as normalized magnetic energy and the overall structure of the magnetic fields might be reliable, although there were remarkable deviations in the absolute value between the instruments.
\cite{2015ApJ...811..107D} investigated the influence of spatial resolution on NLFFF with three kinds of methods similar to  \cite{2009ApJ...696.1780D}.
They showed that the free energy tends to be higher with increasing resolutions and magnetic helicity values vary significantly among different resolutions.

In previous studies above, they showed how different results NLFFF produced.
The questions is whether the completely different solutions are produced or not based on the same bottom boundary.
In other words, how unique is the solution  when focusing on one method?
If completely different solutions exist, how can we obtain the result most consistent to the coronal imaging observations?
Although many previous studies applied NLFFF to solar active regions and showed qualitatively agreement between NLFFF and coronal images \citep{2004A&A...425..345R,2012SoPh..278...73V,2012ApJ...748...77S,2012ApJ...760...17I, 2014ApJ...786L..16J, 2014ApJ...791...84W,2017ApJ...842..106K, 2018ApJ...863..162M}, there still exist some regions where NLFFF produce quite different field lines from coronal images \citep{2009ApJ...696.1780D}.
Therefore, the problem we have to tackle is to reveal whether completely different solution with same bottom boundary can exist or not in the practical calculation and what we should do in case that the modeled field lines do not agree with the coronal images.

To investigate whether completely different solutions with the same bottom boundary exist or not, we evaluate NLFFF extrapolation with different initial conditions.
As is often the case with the nonlinear inverse problems, the different initial guesses may often produce completely different converged solutions.
Therefore the uniqueness of the NLFFF calculation can be studied by giving different initial conditions.

The NLFFF extrapolation is usually performed as follows,
\begin{description}
\item[(1)] Set the 3D initial condition by using photospheric vertical magnetic field.
\item[(2)] Give the information of the horizontal magnetic field to the bottom boundary.
\item[(3)] Perform some relaxation process. 
\end{description}
In the process (1), almost all previous studies use the potential field as an initial condition.
We perform NLFFF calculation not only with the potential field but also with linear force-free field as an initial condition.
Several force-free alpha values are given in the linear force-free case.
Comparisons among the NLFFF results with different initial conditions will provide some insights to the uniqueness of the NLFFF extrapolation.
As described above previous studies show that different results are obtained depending on the calculation methods used in the NLFFF modeling \citep{2006SoPh..235..161S,2008SoPh..247..269M, 2008ApJ...675.1637S,2009ApJ...696.1780D}.
The one of the causes of the difference between the Grad-Rubin method and the other two methods (optimization and MHD relaxation methods) is the treatment of the bottom boundary.
Because the former method relax the 3D magnetic field so that the force-free alpha along the magnetic field line is constant, the horizontal component of magnetic field at the bottom boundary is different from the observational magnetic field in the photosphere.
On the other hand, the latter methods keep the bottom boundary as the observational magnetic field vector in the photosphere.
We give priority to keeping the magnetic field measured in the photosphere at the bottom boundary.
Therefore, we focus on one method, MHD relaxation method, which is developed by \cite{2014ApJ...780..101I}.

The paper is organized as follows: We describe the observations and data reduction in Section \ref{observations}. The method of  the NLFFF extrapolation is described in Section \ref{sec2:mhd_relaxation}.  We present and discuss the results in Sections \ref{sec2:results} and \ref{sec2:discussion}, respectively.
We summarize our results in Section \ref{sec2:summary}.
\section{Observations
\label{observations}}
As targets to analyze, we chose two active regions, NOAA 11692 and NOAA 11967 to investigate how significantly different coronal magnetic field structures are derived depending on initial conditions.
The examined active regions are two extreme examples; one has rather simple bi-pole magnetic distribution at the photosphere and the other shows a complicated field distribution at the photosphere.
The photospheric vertical magnetic field distributions are shown in the left upper and lower panels of Figure \ref{sec2:fig:observation}.
While the former is composed of bi-pole magnetic fields and shows a weak twist in the photosphere, the latter has multi-pole magnetic fields and shows a strong twist.
The global alpha is one of the index values for expressing the degree of the the twist in the active regions.
\cite{2009ApJ...700..199T,2009ApJ...702L.133T} defined global alpha $\alpha_{g}$ such as,
\begin{equation}
\alpha_{g}=\frac{\sum(\frac{\partial B_y}{\partial x}-\frac{\partial B_x}{\partial y})B_z}{\sum B_z^2},
\label{eqn:galpha}
\end{equation}
which can be calculated from the photospheric vector magnetic field, $B_x$, $B_y$, and $B_z$, where z-axis is defined as the vertical direction to the solar surface.
 A global alpha of NOAA 11692 is  $\alpha_{g}=-1.0 \times 10^{-8} {\rm m^{-1}}$ and NOAA 11967 has $\alpha_{g}=-5.0\times 10^{-8} {\rm m^{-1}}$.
Absolute values of $\alpha_{g}$ in various active regions usually vary from $0$ to $5.0 \times 10^{-8} {\rm m^{-1}}$ \citep{1995ApJ...440L.109P, 1997ApJ...481..973P}.
Therefore, the two active regions we analyzed are contrasting examples in terms of the twist. 
\subsection{Observations at NOAA 11692}
\begin{figure}
\includegraphics[bb=0 0 1145 958,scale=0.45]{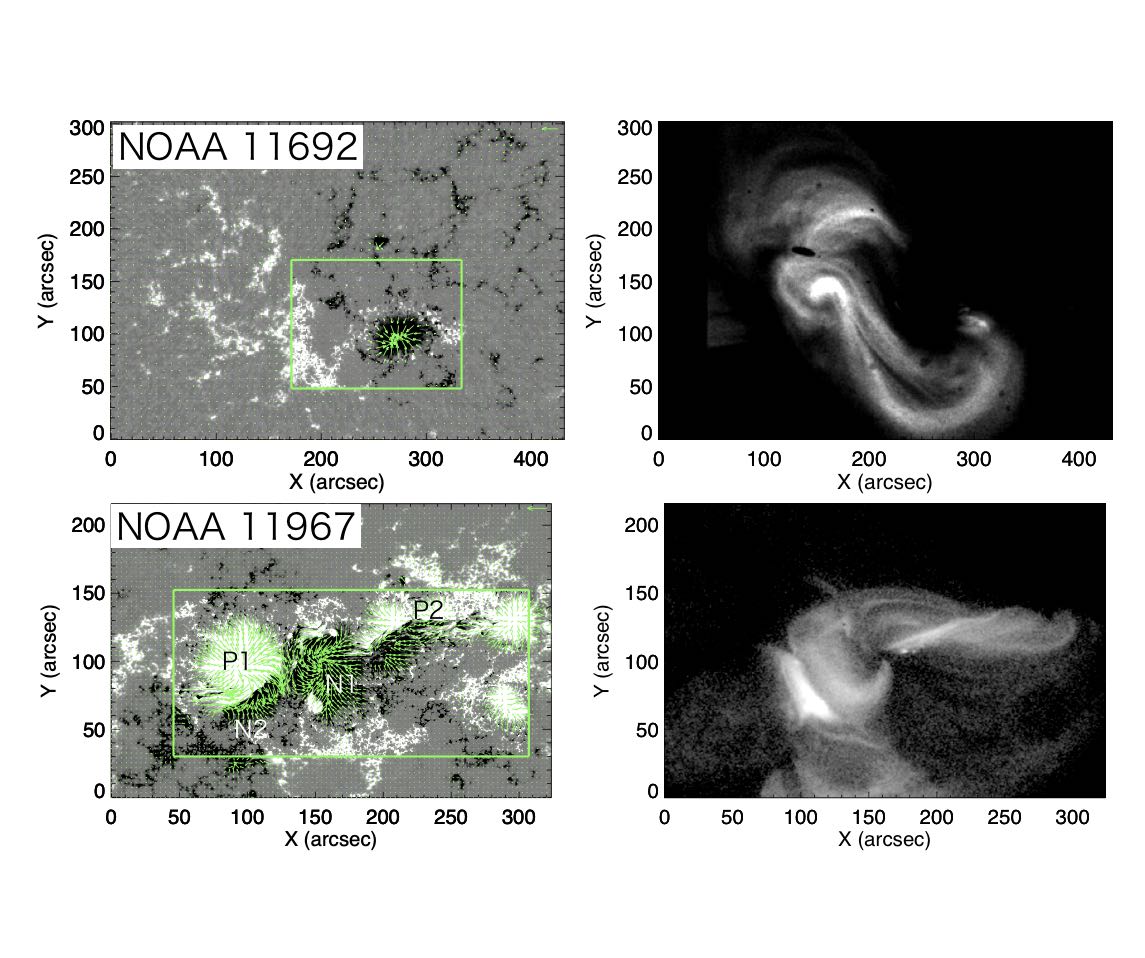}
\caption{Left upper panel: The spatial distribution of the magnetic field vertical to the solar surface in NOAA 11692. {\it Hinode}/SOT SP and {\it SDO}/HMI data are combined in this map. The green rectangle shows the FOV of {\it Hinode}/SOT SP scanned between 03:00 and 03:35 UT on 15 Mar 2013. The green arrows show the horizontal magnetic field.  Right upper panel: Soft X-ray image of NOAA 11692 observed with {\it Hinode}/XRT. Left lower panel: The spatial distribution of the vertical magnetic field in NOAA 11967.  {\it Hinode}/SOT SP and {\it SDO}/HMI data are combined in this map. The green rectangle shows the FOV of {\it Hinode}/SOT SP observed at 07:50-08:45 UT on 3 February 2014. The green arrows show the horizontal magnetic field. Right lower panel:  Soft X-ray image of NOAA 11967 observed with {\it Hinode}/XRT.}
\label{sec2:fig:observation}
\end{figure}

NOAA active region 11692 was a typical active region with a round leading sunspot consisting of an umbra and penumbra.
Several opposite polarity magnetic field are broadly distributed at the following area.
This active region produced 14 C-class flares and 2 M-class flares between 12 Mar 2013 and 22 Mar 2013.
The vector magnetic field map in the upper left of Figure \ref{sec2:fig:observation} was observed with {\it Hinode}/SOT SP and {\it SDO}/HMI.
The FOV of the SP observation is shown by a green box in Figure \ref{sec2:fig:observation}.
To increase the narrow FOV of SP when extrapolating the coronal magnetic field by NLFFF modeling, we used the data obtained by the HMI. 
The gray scale shows the vertical components of the magnetic field to the solar surface and the green arrows show the horizontal magnetic field, which are derived by the inversion of the spectropolarimetric data described  in Section \ref{sec2:observation:reduction}.
The leading spot has the negative polarity and the positive polarity is dominant at the following region.
The negative spot has the anti-clockwise horizontal magnetic field while the strong horizontal magnetic field can not be identified  in the positive following region.
The SP performs spectropolarimetric observations with  two magnetically sensitive Fe {\sc i} lines at 6301.5 \AA \ and 6302.5 \AA \   with a spectral sampling of 21.5 m\AA \ per pixel and scanned this region between 03:00 and 03:35 UT on  15 Mar 2013.
The map has an effective pixel size of 0$''$.3 with the FOV of 166$''$ $\times$ 123$''$.  
Although the SP provides the polarimetric information based on the slit observations with the high spectral sampling of 21.549 m\AA \ pix$^{-1}$,  the slit observations limit the field of view (FOV).
NOAA 11692 was located around disk center (-161$''$, 257$''$) at the time of the SP scanning.
The HMI measures polarization  based on the 6 narrow bands (band width: 76 m\AA \ +/- 10 m\AA \ ) observations around Fe {\sc i} 6173 \AA \ line.
The HMI has an advantage of the regular observation of the full disk of  the Sun with the spatial sampling of 0.$''$5 pix$^{-1}$.
We also used the HMI data obtained at 03:11 UT on 15 Mar 2013.

To evaluate the validity in the result from the NLFFF extrapolations, we utilized the soft X-ray image observed with the X-ray telescope \citep[XRT:][]{2007SoPh..243...63G} on board {\it Hinode}, as shown in the upper right panel of Figure \ref{sec2:fig:observation}.
The image was observed with Be-thin filter at 03:57 UT on 15 Mar 2013 with a field of view of 395$''\times$395$''$. 
The pixel sampling is 1.0$''$.
A sigmoidal structure can be clearly identified in the X-ray image.

\subsection{Observations at NOAA 11967}
NOAA active region 11967 was a flare-productive active region, which produced 10 M-class and 38 C-class flares from 31 January 2014 to 9 February 2014. 
In the same way as the data for NOAA 11692, we combined a magnetic field map of SP and HMI.
We used one of  the SP scanned maps obtained  at 07:50-08:45 UT on 3 February 2014, as shown in the green box in the lower left panel of Figure \ref{sec2:fig:observation}.
The gray scale shows the vertical components of the magnetic field to the solar surface and the green arrows show the horizontal magnetic field, which are derived by the inversion of the spectropolarimetric data described  in Section \ref{sec2:observation:reduction}.
The map has an effective pixel size of 0$''$.3 with a field of view (FOV) of 280$''\times$130$''$.  
NOAA 11967 was located at almost disk center (-100$''$, -100$''$) at the time of the SP scanning.
We used the HMI data obtained at 07:47 UT.
NOAA 11967 was mainly composed of four magnetic polarities (P1, N1, P2, and N2 in Figure \ref{sec2:fig:observation}).
While P1 shows round-shape structure, N1, N2, and P2 show the elongated structures.
At the polarity inversion line (PIL) between N1 and P2, the sheared horizontal magnetic fields are well visible.

We also used X-ray images obtained with XRT with Be-thin filter at 07:12 UT on 3 February 2014 with a field of view of 512$''\times$512$''$, as shown in the lower right panel of Figure \ref{sec2:fig:observation}. 
Compared to NOAA 11692, NOAA 11967 had a complex structure of coronal loops, because this active region has multiple locations of the both polarities.
The sheared magnetic field lines between N1 and P2 also can be seen in the X-ray image.

\subsection{Data Reduction
\label{sec2:observation:reduction}}
For the calibration of the Stokes profiles obtained with {\it Hinode}/SOT SP, we used the Solarsoft routine SP\_PREP \citep{2013SoPh..283..601L}  and applied  a {\it Milne-Eddington atmosphere} (ME) model in order to derive the physical parameters by a nonlinear least square fitting using the code based on MELANIE \citep{2001ASPC..236..487S}. 
 SP\_PREP corrects the wrap-around of the Stokes I, dark and flat, instrumental polarization, spectral line curvature, thermal drift, and orbital Doppler shift.
When we derive the magnetic field azimuth, there is well-known ambiguity called 180 degree ambiguity in the LOS reference frame \citep{2004ASSL..307.....L}.
The 180 degree ambiguity in the transverse magnetic field direction was solved with the minimum energy ambiguity resolution method \citep{1994SoPh..155..235M,2009ASPC..415..365L}.

For the data of HMI, the vector magnetic field data products are provided by the HMI team, which is called  Space-weather HMI Active Region Patches \citep[SHARP;][]{2014SoPh..289.3549B}. The HMI data was used to expand the FOV of the 4$\times$4 binned data of SP ($1.2$''$/$pix) for NOAA 11692 and the 2$\times$2 binned data of SP ($0.6$''$/$pix) for NOAA 11967.
The binning was performed in order to reduce the calculation time of the NLFFF extrapolation.
The spatial resolution affect the energy and free energy of NLFFF modelings \citep{2015ApJ...811..107D}, because the binning process changes the magnetic energy and free energy at the bottom boundary.
In this study, we focus on the dependence of the initial condition on the results of the NLFFF based on the same bottom boundary.
Therefore, we perform the NLFFF modeling based on the bottom boundary with single binning factor for each active region ($1.2$''$/$pix for NOAA 11692 and $0.6$''$/$pix for NOAA 11967).

\section{MHD Relaxation Method and Numerical Settings
\label{sec2:mhd_relaxation}}
The nonlinear force-free field extrapolation is performed by the MHD relaxation method \citep{2014ApJ...780..101I,2016PEPS....3...19I}, which uses the following equations,
\begin{eqnarray}
\frac{\partial \vector{v}}{\partial t}&=&-(\vector{v}\cdot \vector{\nabla})\vector{v}+\frac{1}{\rho}\vector{j}\times\vector{B}+\nu\nabla^2\vector{v}, \label{momentum} \\
\frac{\partial \vector{B}}{\partial t}&=&\nabla \times (\vector{v}\times \vector{B}-\eta \vector{j})-\nabla \phi, \label{induction}\\
\vector{j}&=& \nabla \times \vector{B}, \label{ampere}\\
\frac{\partial \phi}{\partial t} &+&c_{h}^2\nabla\cdot \vector{B}=-\frac{c_{h}^2}{c_{p}^2}\phi, \label{9wave}
\end{eqnarray}
where $\rho$ is the pseudo density, which is assumed to be equal to $|\vector{B}|$ to ease the relaxation by equalizing the Alfv\'en speed in space,  $\phi$ is the convenient potential for $\nabla \cdot \vector{B}$ cleaning and $\nu$ is the viscosity, which is set to a constant ($1.0 \times 10^{-3}$). The length, magnetic field, velocity, and time were normalized by $L_{0}=157 $ Mm for NOAA 11967 and $L_{0}=314$ Mm for NOAA 11692 and $B_{0}=4000$ G, $V_{\rm A} \equiv B_0/(4\pi \rho_0)^{1/2}$, and $\tau_{A} \equiv L_{0}/V_{A}$, where $V_{A}$ is the Alfv$\rm {\acute{e}}$n velocity. Equations (\ref{momentum}), (\ref{induction}), (\ref{ampere}) and  (\ref{9wave}) are the equation of motion,  the induction equation, the ${\rm Amp\grave{e}re's}$ law, and $\nabla\cdot\vector{B}$ cleaning introduced by \cite{2002JCoPh.175..645D}, respectively. The parameters $c_{p}^2$ and $c_{h}^2$ are the advection and diffusion coefficients, respectively and fixed at $0.1$ and $0.04$. The non-dimensional resistivity $\eta$ is given by
\begin{equation}
\eta=\eta_{0}+\eta_{1}\frac{|\vector{j}\times\vector{B}||\vector{v}|^2}{|\vector{B}|^2},
\label{resistivity}
\end{equation}
where $\eta_{0}$ and $\eta_{1}$ are fixed at $5.0\times 10^{-5}$ and $1.0\times10^{-3}$ in non-dimensional units.  
The second term is introduced to accelerate the relaxation to the force-free state.

The velocity field at each grid was adjusted at each time step below in order to avoid becoming large value.
When the value of $v^{*}$ becomes larger than the value of $v_{\rm{max}}$,
\begin{equation}
\vector{v} \rightarrow \frac{v_{{\rm max}}}{v^{*}}\vector{v},
\end{equation}
where $v^{*}=|\vector{v}|/|\vector{v_{A}}|$, and $v_{{\rm max}}=0.1$. 

In previous NLFFF calculations, the potential field has been used as initial guess for the 3D magnetic field structure. 
 We chose linear force free field as an initial condition including a potential field, which satisfies $\nabla\times \vector{B}=\alpha_0 \vector{B}$, where $\alpha_0$ is  a constant.
We examined different 5 initial conditions  $\alpha_0=[0, \pm 1.2, \pm 2.3] \ \times 10^{-8} {\rm m^{-1}}$ for NOAA 11692 and 12 initial conditions $\alpha_0=[0, \pm 0.70, \pm 1.2, \pm 2.3, \pm 4.6, \pm 7.0, -12] \ \times 10^{-8} {\rm m^{-1}}$ for NOAA11967.
The values of $\alpha_0$ are chosen by the pixel size. 
The $\alpha_0$ values of $[0, \pm0.70, \pm1.2, \pm2.3, \pm4.6, \pm7.0, -12] \times 10^{-8}$ $m^{-1}$ correspond to $[0, \pm0.003, \pm0.005, \pm0.01, \pm0.02, \pm0.03, -0.05] $ pix$^{-1}$ for NOAA11967. Because the magnetic field of NOAA 11967 is less affected by the initial condition than that of NOAA 11692, we have more initial conditions for NOAA 11967.

\begin{figure}
\includegraphics[bb=0 0 504 600]{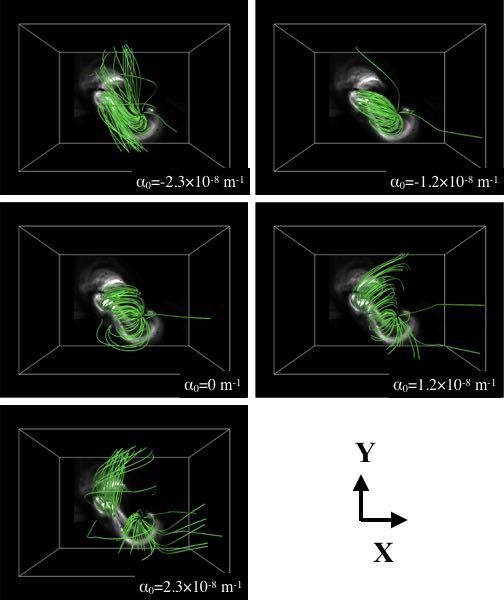}
\caption{The morphology of magnetic field lines (green solid lines) in NOAA 11692 at the initial condition for the NLFFF extrapolation. The background grayscale image is soft X-ray images observed with {\it Hinode}/XRT.}
\label{sec2:initial11692}
\end{figure}

\begin{figure}
\includegraphics[bb=0 0 675 626,scale=0.7]{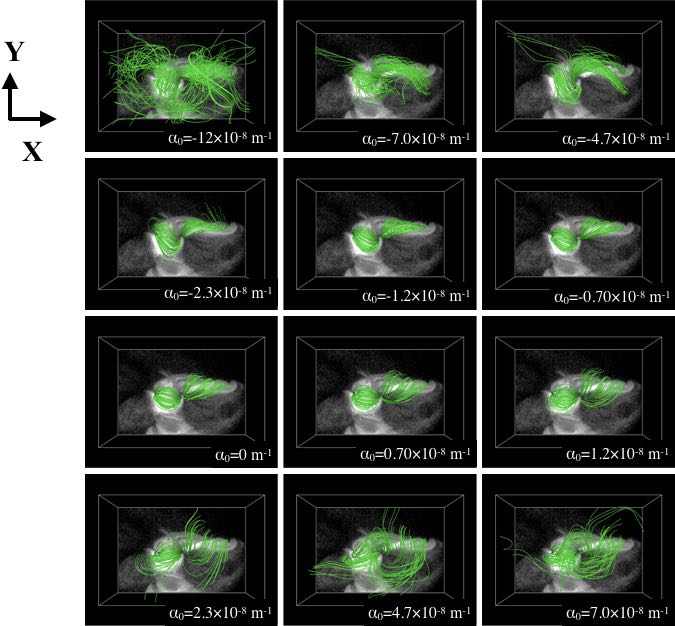}
\caption{The morphology of magnetic field lines (green solid lines) in NOAA 11967 at the initial condition for the NLFFF extrapolation.The background grayscale image is soft X-ray images observed with {\it Hinode}/XRT.}
\label{sec2:initial11967}
\end{figure}

We used the equations of \cite{1981A&A...100..197A} for the calculations of the linear force-free field and the resulting initial conditions are shown in Figures \ref{sec2:initial11692} and \ref{sec2:initial11967}. 
Although the case $\alpha_0=+12 \ \times 10^{-8} {\rm m^{-1}}$ for NOAA 11967 was also calculated, the calculation did not converge.
Therefore, we do not include it in the results in this paper.
The numerical domain is set to $(0,0,0) <(x,y,z)<(1.0,0.7,0.7)$ resolved by $360\times252\times252$ nodes for NOAA11692 and $(0,0,0) <(x,y,z)<(1.5,1.0,0.5)$ resolved by $540\times360\times180$ nodes for NOAA 11967. 
In order to set the same top boundary for all  calculations, the initial conditions above $z=0.417$ are set to the potential field.

The magnetic field at the top was fixed with the initial state (potential field) and the normal component of the magnetic field on the bottom boundary was also fixed. The side boundary was periodic.
We varied the transverse component on the bottom boundary $\vector{B}_{\rm BC}$ as follows,
\begin{equation}
\vector{B}_{\rm BC}=\gamma\vector{B}_{\rm obs}+(1-\gamma)\vector{B}_{\rm intial},
\label{bottombc}
\end{equation}
where $\vector{B}_{\rm obs}$ and $\vector{B}_{\rm initial}$ are the transverse component of the observational and initial bottom boundary, respectively. 
We increased $\gamma=\gamma+d\gamma$ when $\int |\vector{j}\times\vector{B}|^2dV$ dropped below a critical value. In this study we set $d\gamma=0.1$.
When $\gamma$ becomes equal to $1$, $\vector{B}_{BC}$ is consistent with the observed field.
The number of calculation steps were set  to 24000 steps for NOAA 11692 and 25000 steps for NOAA 11967

Spatial derivatives are calculated by the second-order central differences and  temporal derivatives are integrated by the Runge-Kutta-Gill method to fourth order accuracy. 

\section{Results
\label{sec2:results}}
\subsection{Properties of the Active Regions}
\begin{figure}
\includegraphics[bb=0 0 1133 566,scale=0.4]{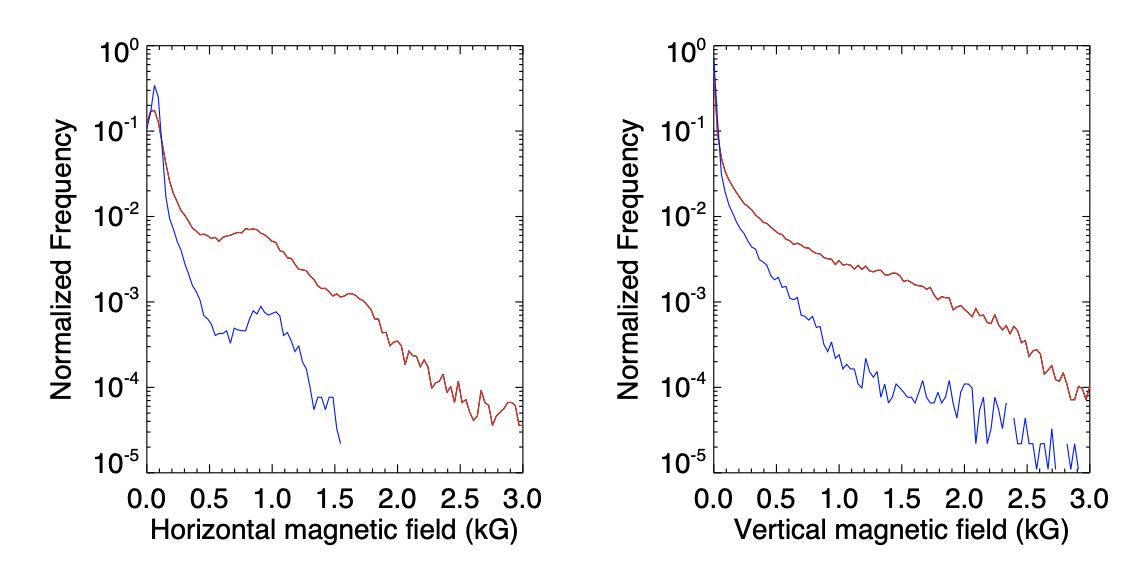}
\caption{The histogram of the horizontal magnetic field (left panel) and vertical magnetic field (right panel) for NOAA 11692 (blue solid line) and NOAA 11967 (red solid line)}
\label{sec2:histo_hor_vertical}
\end{figure}

In this Section, the property of two active regions in the photosphere are summarized.
Figure \ref{sec2:histo_hor_vertical} shows the histogram of the horizontal magnetic field (left panel) and vertical magnetic field (right panel) for NOAA 11692 (blue solid line) and NOAA 11967 (red solid line).  
For both horizontal and vector magnetic field, NOAA 11967 has larger frequency at larger magnetic field.
While 6.3 $\%$ of the horizontal magnetic field in the FOV is larger than 1000 G for NOAA 11967, 0.49 $\%$ is larger than 1000 G for NOAA 11692.
Regarding the vertical magnetic field, the ratios of $B_z > 1000$G are 7.5 $\%$ and 0.47 $\%$ for NOAA 11967 and NOAA 11692, respectively. 
The  unsigned vertical magnetic flux of NOAA 11967 is $8.4\times10^{22}$ Mx, which is 2.3 times larger than that of NOAA 11692, $3.7\times 10^{22}$ Mx.


 One issue with performing NLFFF extrapolations from the photosphere is that this layer is considered to contain Lorentz forces \citep[e.g.,][]{2001SoPh..203...71G}.
There are some previous studies to investigate the force-freeness in the active regions in the photosphere based on the necessary condition of the force-free approximation shown by \cite{1985svmf.nasa...49L}.
The Lorentz force can be written as the divergence of the Maxwell stress tensor, 
\begin{equation}
M_{ij}=-\frac{B^2}{8\pi} \delta_{ij}+\frac{B_{i}B_{j}}{4\pi}.
\end{equation}
 Assuming that the strength of magnetic field vanishes at infinite height, the volume-integrated Lorentz force can be written by the surface integrals,
\begin{eqnarray}
F_x&=&\frac{1}{4\pi}\int B_{x}B_{z}dxdy , \label{fx}\\
F_y&=&\frac{1}{4\pi}\int B_{y}B_{z}dxdy ,  \label{fy}\\
F_z&=&\frac{1}{8\pi}\int (B_{z}^2-B_{x}^2-B_{y}^2)dxdy.  \label{fz}
\end{eqnarray}
According to \cite{1985svmf.nasa...49L}, in order to regard the magnetic field as being force-free state, three components of the net Lorentz force are necessary to sufficiently be smaller than the integrated magnetic pressure force,
\begin{equation}
F_{p}=\frac{1}{8\pi}\int (B_{x}^2+B_{y}^2+B_{z}^2)dxdy. \label{fp}
\end{equation}
The force-freeness of the active regions from Equations (\ref{fx}), (\ref{fy}), and (\ref{fz}) are listed in Table \ref{sec2:table:forcefreeness}.
The values of $|F_x|/F_p$, $|F_y|/F_p$, and $|F_z|/F_p$ of NOAA 11967 are sufficiently small, i.e., the active region satisfies the necessary condition of the force-free field.
On the other hand , the absolute value of the force-freeness of NOAA 11692 is slightly larger than 0.1 in $|F_y|/F_p$ and $|F_z|/F_p$.
However, this value is not so large and not far from 0.1 compared to other active regions reported in the previous studies \citep{1995ApJ...439..474M, 2002ApJ...568..422M, 2013PASA...30....5L}.  
Therefore, the two active region NOAA 11692 and NOAA 11967 are comparatively  appropriate regions applying to the assumption of the force-free modeling.
  \begin{table}[H]
\centering
\caption{Force-freeness of the active regions derived from Eqns (\ref{fx}), (\ref{fy}), and (\ref{fz})}
 \begin{tabular}{cccc} 
\hline
  & NOAA 11692  & NOAA 11967  \\
\hline
 $|F_x|/F_p$& 0.053&0.029 \\
  $|F_y|/F_p$& 0.128&0.0089 \\
   $|F_z|/F_p$& 0.17&0.089 \\
\hline
\end{tabular}
\label{sec2:table:forcefreeness}
\end{table}

\subsection{Morphology of Field Lines from NLFFF
\label{sec2:morphology}}
\begin{figure}
\includegraphics[bb=0 0 502 600]{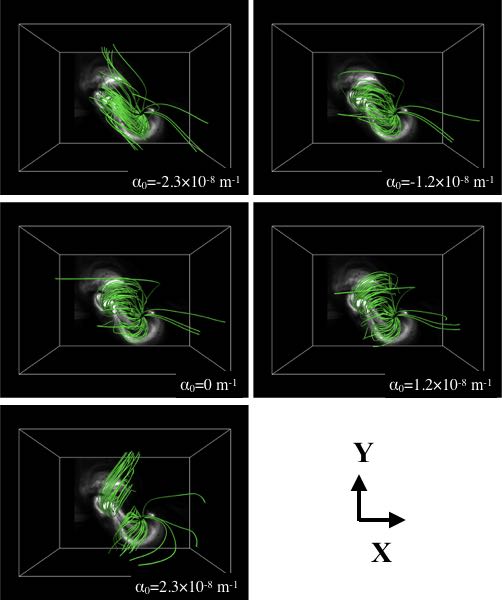}
\caption{The morphology of magnetic field lines (green solid lines) in NOAA 11692 as a result of the NLFFF extrapolation. The background grayscale image is soft X-ray images observed with {\it Hinode}/XRT.}
\label{sec2:nlfff11692}
\end{figure}

\begin{figure}
\includegraphics[bb=0 0 1012 326,scale=0.45]{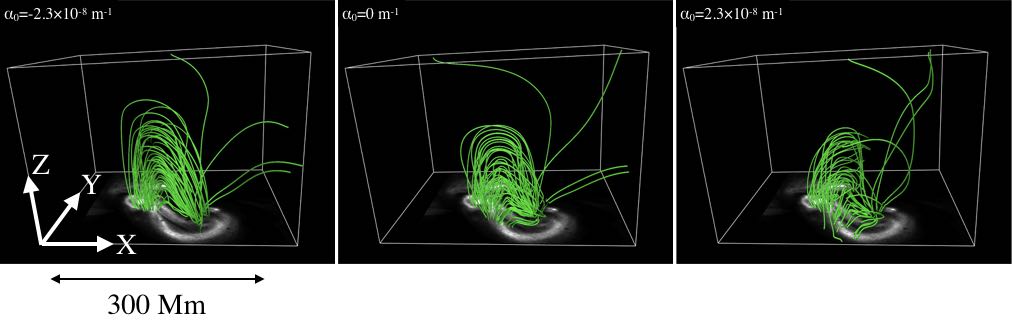}
\caption{A side view of the magnetic field lines in NOAA 11692, obtained by the NLFFF extrapolation with three initial conditions. The background grayscale image projected on the bottom surface are a soft X-ray image observed with {\it Hinode}/XRT.}
\label{sec2:nlfff11692_difangle}
\end{figure}

Figure \ref{sec2:nlfff11692} shows the magnetic field lines (green solid lines) in NOAA 11692 as a result of the NLFFF extrapolation with 5 different initial conditions.
The magnetic field lines are chosen randomly around the region where the sigmoidal structure is identified.
Back ground gray scale images are X-ray images observed with the XRT.
As clearly seen, the 3D morphology of the magnetic field lines strongly depends on the initial condition.
In other words, the morphology of  magnetic field lines from the NLFFF extrapolation is not far from that of the initial condition.
When we use the potential field as an initial condition ($\alpha_0=0$ case), which is usually chosen, the field lines are potential-like and the sigmoidal structure can not be correctly reproduced.
On the other hand, when we choose appropriate initial condition (e.g. $\alpha_0=-2.3\times10^{-8}$ m$^{-1}$ in Figure \ref{sec2:nlfff11692}), many magnetic field lines from the NLFFF extrapolation are almost parallel to the direction of the sigmoidal structure in the X-ray image.
The value  $\alpha_0=-2.3\times10^{-8}$ m$^{-1}$  is larger than the global alpha estimated from the photospheric magnetic field, $\alpha_g=-1.0\times10^{-8}$ m$^{-1}$.
Figure \ref{sec2:nlfff11692_difangle} shows a side view of the magnetic field lines.
Only the cases of $\alpha_0=-2.3\times10^{-8}$, $0$, and $2.3\times10^{-8}$ m$^{-1}$ are shown.
The height of the field lines are at around $100-200$ Mm.
Not only long loop lines whose loop top is located at more than 100 Mm, the short field lines whose loop top is located below 100 Mm are also different among different solutions.

\begin{figure}
\includegraphics[bb=0 0 676 626,scale=0.75]{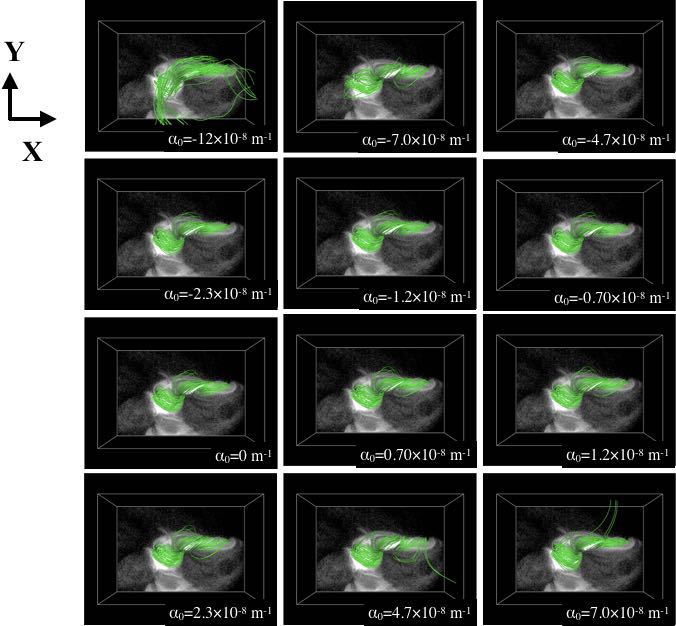}
\caption{The morphology of magnetic field lines (green solid lines) in NOAA 11967 as a result of the NLFFF calculation. The background grayscale image is a soft X-ray image observed with {\it Hinode}/XRT.}
\label{sec2:nlfff11967}
\end{figure}

\begin{figure}
\includegraphics[bb=0 0 1011 326,scale=0.45]{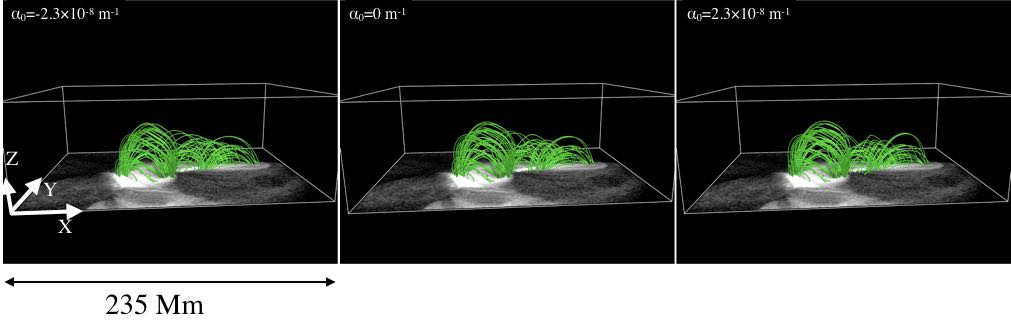}
\caption{A side view of the magnetic field lines in NOAA 11967, obtained by the NLFFF extrapolation with three initial conditions. The background grayscale image projected on the bottom surface are a soft X-ray image observed with {\it Hinode}/XRT.}
\label{sec2:nlfff11967_difangle}
\end{figure}

Figure \ref{sec2:nlfff11967} shows the magnetic field lines (green solid lines) in NOAA 11967 as a result of the NLFFF extrapolation with 12 different initial conditions.
Compared to NOAA 11692, the results are less dependent on the initial condition.
When we use large $|\alpha_0|$($>7.0\times 10^{-8}$ m$^{-1}$) as an initial condition, the result shows a bit different morphology.
The result of $\alpha_0=-1.2\times 10^{-8}$ m$^{-1}$  does not seem to converge to the reasonable solution.
As shown in Figure \ref{sec2:nlfff11967_difangle}, the height of the field lines is around 30 Mm.

\subsection{Total Magnetic Energy, Total Free Energy, and Extrapolation Metrics}
To evaluate the difference due to the initial condition quantitatively, we focus on the total magnetic energy, total free energy and extrapolation metrics, as shown in Table \ref{sec2:table:metrics_11692} for NOAA 11692 and Table \ref{sec2:table:metrics_11967} for NOAA 11967.
The first column shows the constant force-free alpha $\alpha_0$, which is used for the initial condition of the NLFFF extrapolation.
The second, third and fourth columns show the magnetic energy of the initial condition $E_{\rm init}$, the magnetic energy of the NLFFF $E$, and the free magnetic energy $E_{\rm free}$.
The third and fourth columns are normalized by the magnetic energy of the potential field $E_{\rm pot}$.
For NOAA 11692, when we choose the potential field as an initial condition, the resulting magnetic energy of the NLFFF becomes almost potential.
 When the magnetic energy of the initial condition is larger, the resulting magnetic energy of the NLFFF becomes larger.
However the magnitude of the difference of $E/E_{\rm pot}$ among the different initial condition is not so large compared to that of $E_{\rm init}$.
The ratios of maximum total magnetic energy and free energy ($\alpha_0=-2.3\times 10^{-8}$ m$^{-1}$) to minimum ones ($\alpha_0=0$ m$^{-1}$) are 1.08 and 2.64, respectively.
NOAA 11967 has larger magnetic energy and free energy than NOAA 11692 does as shown in Tables \ref{sec2:table:metrics_11692} and \ref{sec2:table:metrics_11967}.
As mentioned in Section \ref{sec2:morphology}, the result from $\alpha_0=-12\times10^{-8}$ m$^{-1}$ does not seem to be physically reasonable.
Therefore we focus on the results except $\alpha_0=-12\times10^{-8}$ m$^{-1}$ for NOAA 11967.
Even when we choose the potential field as an initial condition, the resulting NLFFF has $E_{\rm free}/E_{\rm pot}=0.14$.
The notable result for NOAA 11967 is in spite of the large initial constant $\alpha_0$ such as $|\alpha_0|>2.3\times10^{-8}$ m$^{-1}$, the resulting magnetic field energy and free energy do not show large difference among the different initial conditions.
The ratios of maximum total magnetic energy and free energy ($\alpha_0=-7.0\times10^{-8}$ m$^{-1}$) to minimum ones ($0$ m$^{-1}$) are 1.03 and 1.28, respectively.

The fifth column shows $\langle{\rm CW}\sin\theta\rangle$, the mean sine of the angle $\theta$ between \vector{j} and \vector{B} weighted by $\vector{j}$, which is defined as follows,

\begin{equation}
\langle{\rm CW}\sin\theta\rangle=\frac{\sum|\vector{j}\sin\theta|}{\sum |\vector{j}|}.
\end{equation}
The metric is based on the property of the force-free field that the electric current is parallel to the magnetic field.
Therefore  $\langle{\rm CW}\sin\theta\rangle$ represents how force-free an obtained NLFFF solution is.
When the solution is close to the force-free state, $\langle{\rm CW}\sin\theta\rangle$ is close to zero.
For NOAA 11692, the $\langle{\rm CW}\sin\theta\rangle$ has the smallest value when $\alpha_0=-2.3 \times10^{-8}$ m$^{-1}$, while it has the largest value when $\alpha_0=0$ m$^{-1}$.
For NOAA 11967, the magnitude of $\langle{\rm CW}\sin\theta\rangle$  is smaller than that of NOAA 11692.
Similar to NOAA 11692, the $\langle{\rm CW}\sin\theta\rangle$ tends to have smaller value when the $|\alpha_0|$ becomes larger. 
 The magnetic field tends to be force-free, where the field strength is large in our NLFFF modeling, as shown in the analysis of Appendix. 
  This means that an increase of the number of pixels with strong magnetic field leads to smaller values of $\langle{\rm CW}\sin\theta\rangle$. When the absolute initial alpha is large, the number of pixels with strong magnetic field is also large.

The sixth column shows the fractional flux ratio $\langle|f_i|\rangle$
\begin{eqnarray}
f_i &=& \frac{\int_{\Delta S} \vector{B}\cdot d\vector{S}}{\int_{\Delta S}|\vector{B}|dS} \nonumber\\
&=&\frac{(\nabla\cdot\vector{B})\times (\Delta x)^3}{|\vector{B}|\times6(\Delta x)^2}\nonumber\\
&=&\frac{(\vector{\nabla\cdot\vector{B})}\Delta x}{6|\vector{B}|},
\end{eqnarray}
where $\Delta S$ and $\Delta x$ are the small discrete surface and the grid spacing. 
From one of the Maxwell equations, the divergence of the magnetic field must vanish in the calculation box.
The  $\langle|f_i|\rangle$ represents how divergence-free an obtained NLFFF solution is.
 When $\vector{\nabla}\cdot\vector{B}$  very nearly vanishes in the results of the NLFFF extrapolation, the $\langle|f_i|\rangle$ becomes close to zero.
For NOAA 11692, the $\langle|f_i|\rangle$ has the smallest value, when the initial condition is $\alpha_0=-2.3 \times10^{-8}$ m$^{-1}$.
As shown in Section \ref{sec2:morphology}, the field lines of  the NLFFF from $\alpha_0=-2.3 \times10^{-8}$ m$^{-1}$ are almost parallel to the direction of the sigmoidal structure in the X-ray image.
On the other hand, for NOAA 11967, while the morphology of the field lines are not so different among the NLFFF results, each solution has the different $\langle|f_i|\rangle$ value.
 The $\langle|f_i|\rangle$ has the smallest value, when the initial condition is $\alpha_0=1.2 \times10^{-8}$ m$^{-1}$ and tends to have large value when the $|\alpha_0|$ is large.
 One of the reasons is the setting of the initial condition. In order to set the same top boundary for all calculations, the initial conditions above z = 0.417 are set to the potential field. If the absolute initial alpha is large, there is strong discontinuity in the initial condition. It might be the reason of large div B. The effect of non-zero $\nabla \cdot \vector{B}$ on the energy can be estimated by deriving the non-solenoidal magnetic field propsoed by \cite{2013A&A...553A..38V}. 
 Although we do not analyze the non-solenoidal magnetic field in this paper, the effect of non-zero $\nabla \cdot \vector{B}$ on energy is a few percent if $\langle|f_i|\rangle < 10^{-5}$ from the results of \cite{2013A&A...553A..38V}.


  \begin{table}
\centering
\caption{NLFFF metrics for NOAA 11692}
 \begin{tabular}{cccccc}
\hline
Initial $\alpha_0$ [$10^{-8}$ m$^{-1}$] &$E_{\rm init}$ [$10^{32}$ erg]&$E/E_{\rm pot}$& $E_{\rm free}/E_{\rm pot}$ &$\langle$CWsin$\theta\rangle$&$\langle|f_i|\rangle $[$\times 10^{-5}$]\\
\hline
-2.3 & 8.12&1.14&0.14&0.38&5.58\\
-1.2 &6.46&1.06&0.06&0.54&7.36\\
0 (potential) &6.27&1.05&0.05&0.58&7.90\\
1.2 &6.54&1.06&0.06&0.57&8.09\\
2.3 &8.39&1.14&0.14&0.39&6.35\\
\hline
\end{tabular}
\begin{flushleft}
{\footnotesize First column: The constant force-free alpha $\alpha_0$ used for the initial condition of the NLFFF extrapolation. Second column:The magnetic energy of the initial condition $E_{\rm init}$. Third column:the magnetic energy of the NLFFF $E$. Fourth column: The free magnetic energy $E_{\rm free}$. Fifth Column: The mean sine of the angle $\theta$ between \vector{j} and \vector{B} weighted by $\vector{j}$, which represents force-freeness of the NLFFF extrapolation. Sixth Column: The fractional flux ratio, which represents the divergence-freeness of the NLFFF extrapolation.}
\end{flushleft}
\label{sec2:table:metrics_11692}
\end{table}

 \begin{table}
\centering
\caption{NLFFF metrics for NOAA 11967}
 \begin{tabular}{cccccc}
\hline
Initial  $\alpha_0$ [$10^{-8}$ m$^{-1}$]    &$E_{\rm init}$ [$10^{33}$ erg]&$E/E_{\rm pot}$& $E_{\rm free}/E_{\rm pot}$ &$\langle$CWsin$\theta\rangle$&$\langle|f_i|\rangle $[$\times 10^{-5}$]\\
\hline 
-12 &20.0&1.31&0.31&0.19&8.26\\
-7.0&7.17& 1.18&0.18&0.23&7.60\\
-4.7 &4.41&1.16&0.16&0.25&5.82\\
-2.3 &3.20&1.14&0.14&0.25&7.42\\
-1.2 &3.05& 1.14&0.14&0.26&5.66\\
-0.70 &3.03& 1.14&0.14&0.27&5.16\\
0 (potential)&3.02 &1.14&0.14&0.27&4.50\\
0.70 &3.03& 1.14&0.14&0.27&4.07\\
1.2 &3.06&1.14&0.14&0.27&3.97\\
2.3 &3.22&1.14&0.14&0.27&5.07\\
4.7 &4.38&1.15&0.15&0.26&5.89\\
7.0 &7.19&1.17&0.17&0.24&6.16\\
\hline
\end{tabular}
\begin{flushleft}
{\footnotesize The definition of each column is the same as Table \ref{sec2:table:metrics_11967}.}
\end{flushleft}
\label{sec2:table:metrics_11967}
\end{table}

\subsection{Comparison of Solutions at Each Height
\label{sec2:comaprison_height}}

\begin{figure}
\includegraphics[bb=0 0 985 441,scale=0.5]{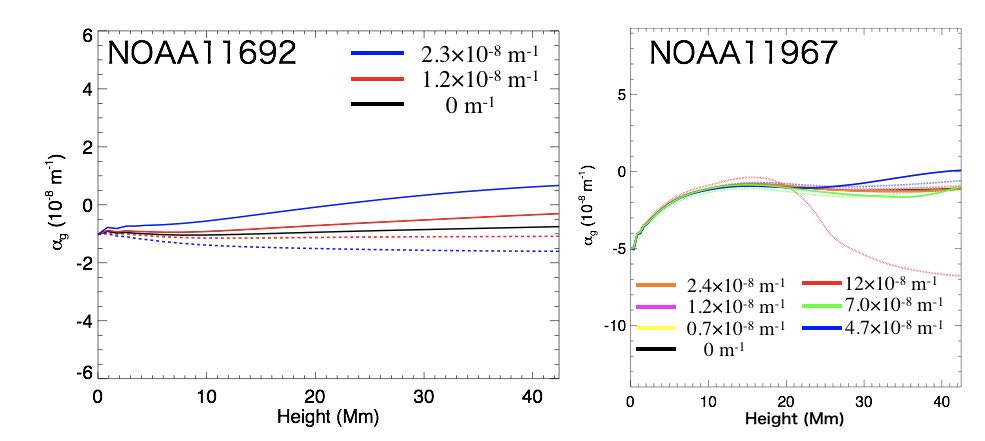}
\caption{Height variation of global alpha. Each color shows each initial condition and solid (dotted) lines show positive (negative) $\alpha_0$.}
\label{sec2:galpha}
\end{figure}

Figure \ref{sec2:galpha} shows the global alpha derived at each height from the NLFFF results.
The global alpha was calculated by Equation (\ref{eqn:galpha}).
Different color shows the results from different initial condition.
For both NOAA 11692 and NOAA 11967,  the global alpha shows larger deviation among the calculations at the higher layer than at the lower.
For NOAA 11692, the global alpha is $-1.0\times 10^{-8}$ m$^{-1}$ at the photospheric height and depending on the initial condition,   the global alpha diverges to the positive and negative values as the height increases.
For NOAA 11967, the global alpha is $-5.0\times 10^{-8}$ m$^{-1}$ at the photospheric height.
Independent of the initial condition, the global alpha increases as the height increases and beyond 20 Mm, the global alpha starts to show the deviation depending on the initial condition.

\begin{figure}
\includegraphics[bb=0 0 850 1133,scale=0.5]{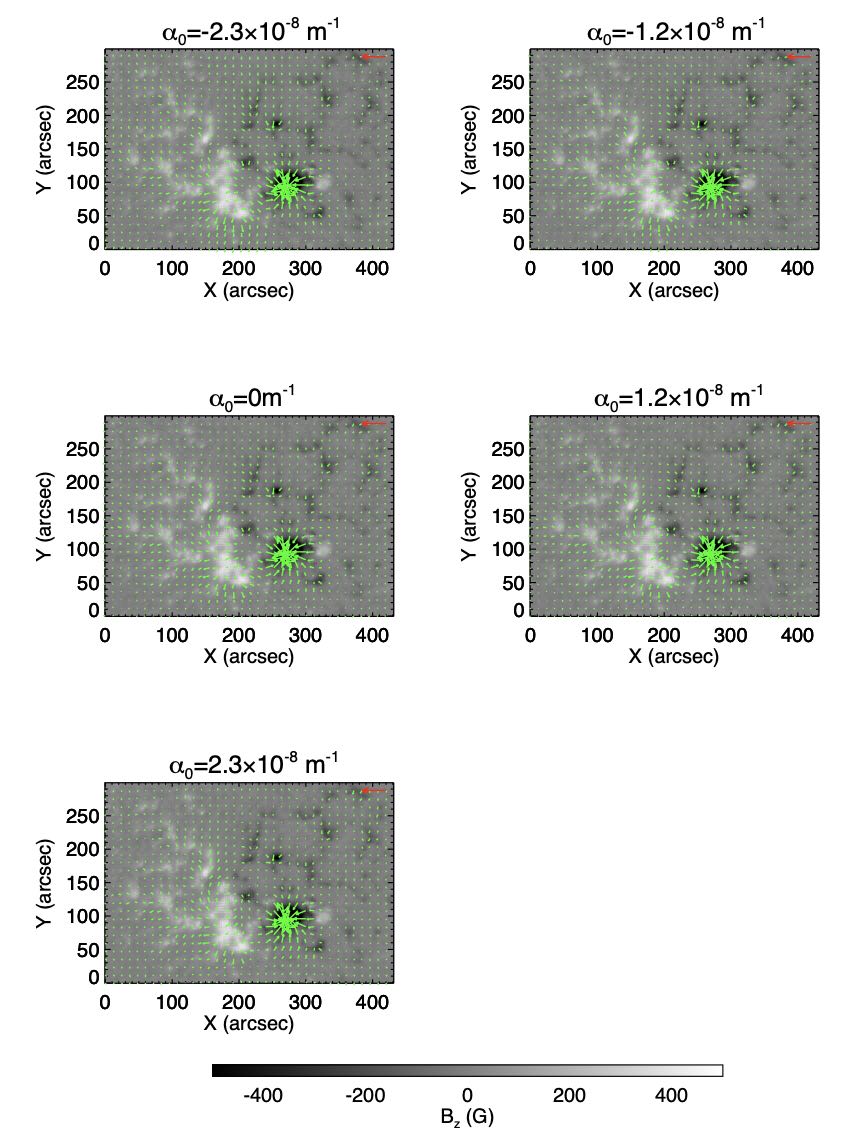}
\caption{The spatial distributions of vector magnetic field for each solution in NOAA 11692 at 2.6 Mm height. The grayscale color shows the vertical magnetic field and green arrows show the horizontal magnetic field. The length of the red arrow shows the field strength of 1000 G.}
\label{sec2:bxbybz11692_3pix}
\end{figure}

\begin{figure}
\includegraphics[bb=0 0 850 1133,scale=0.5]{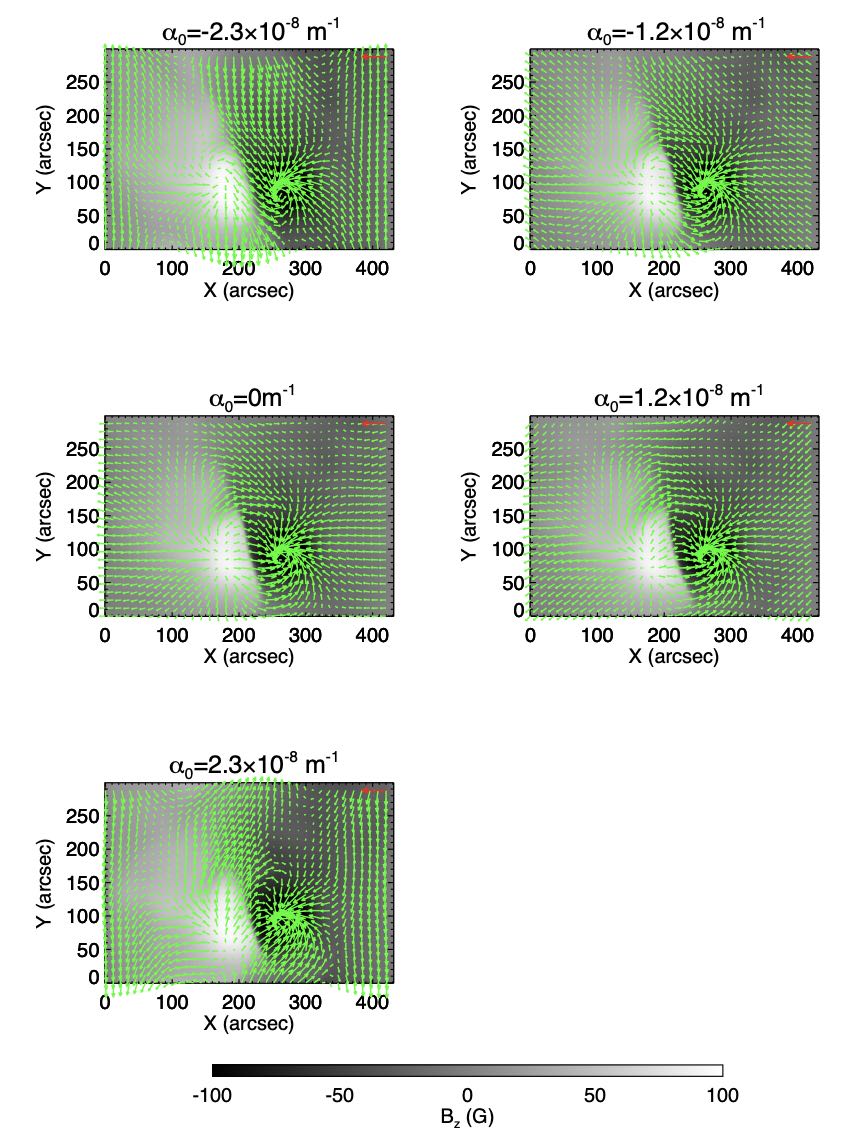}
\caption{Similar as Figure \ref{sec2:bxbybz11692_3pix}, but at 26 Mm. The length of the red arrow shows the field strength of 200 G.}
\label{sec2:bxbybz11692_30pix}
\end{figure}

\begin{figure}
\includegraphics[bb=0 0 800 579,scale=0.6]{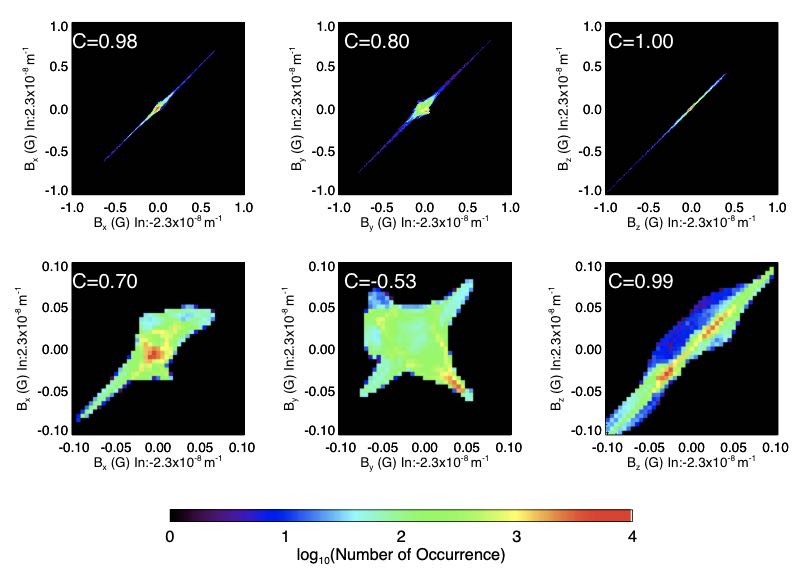}
\caption{All panels show  the number density distribution in vector magnetic field from different initial conditions at 2.6 Mm (upper) and 26 Mm (lower) height. Comparisons between $\alpha_0=0 \ {\rm m^{-1}}$ and $\alpha_0=-2.3\times10^{-8} \ {\rm m^{-1}}$ in NOAA 11692. }
\label{sec2:bxbybz_2dhist_11692}
\end{figure}

The spatial distributions of vector magnetic field in NOAA 11692 at 2.6 Mm and 26 Mm height are shown in Figures \ref{sec2:bxbybz11692_3pix} and \ref{sec2:bxbybz11692_30pix}.
The grayscale color shows the vertical magnetic field and green arrows show the horizontal magnetic field.
The horizontal magnetic field in the negative spot is less dependent on the initial condition at 2.6 Mm.
In the region between positive and negative polarities, however, the horizontal magnetic field is a bit different between results from the different initial conditions.
This tendency can be clearly seen at the higher height as shown in Figure \ref{sec2:bxbybz11692_30pix}.
The horizontal magnetic field around the polarity inversion line in case of $\alpha_0=2.3\times10^{-8}$ m$^{-1}$ is completely different from that in case of $\alpha_0=-2.3\times10^{-8}$ m$^{-1}$.
On the other hand, the vertical magnetic field distribution shows the simple two polarity configuration and is similar to each other.
Figure \ref{sec2:bxbybz_2dhist_11692} shows  the number density distribution in vector magnetic field with different initial condition for NOAA 11692 at the height of 2.6 Mm and 26 Mm.
The result with $\alpha_0=0 \ {\rm m^{-1}}$ is compared with that with $\alpha_0=-2.3\times10^{-8} \ {\rm m^{-1}}$.
As seen in the difference between Figures \ref{sec2:bxbybz11692_3pix} and  \ref{sec2:bxbybz11692_30pix}, the dispersion is larger in 26 Mm compared to in 2.6 Mm.
While  the correlation coefficients, $C$ are  0.96 and 0.95 for $B_x$ and $B_y$ at 2.6 Mm, respectively, those at 26 Mm are 0.82 and 0.60, respectively.
The dispersion also can be seen at 2.6 Mm in the regions with small magnetic field ($\sim 100 $G).
In the number density distribution of $B_y$, some pixels have large $B_y$ in $\alpha_0=-2.3\times10^{-8} \ {\rm m^{-1}}$, while have small $B_y$ in $\alpha_0=0 \ {\rm m^{-1}}$
This distribution reflects the horizontal magnetic field distribution around the polarity inversion line, as described above.

%

\begin{figure}
\includegraphics[bb=0 0 1133 1133,scale=0.45]{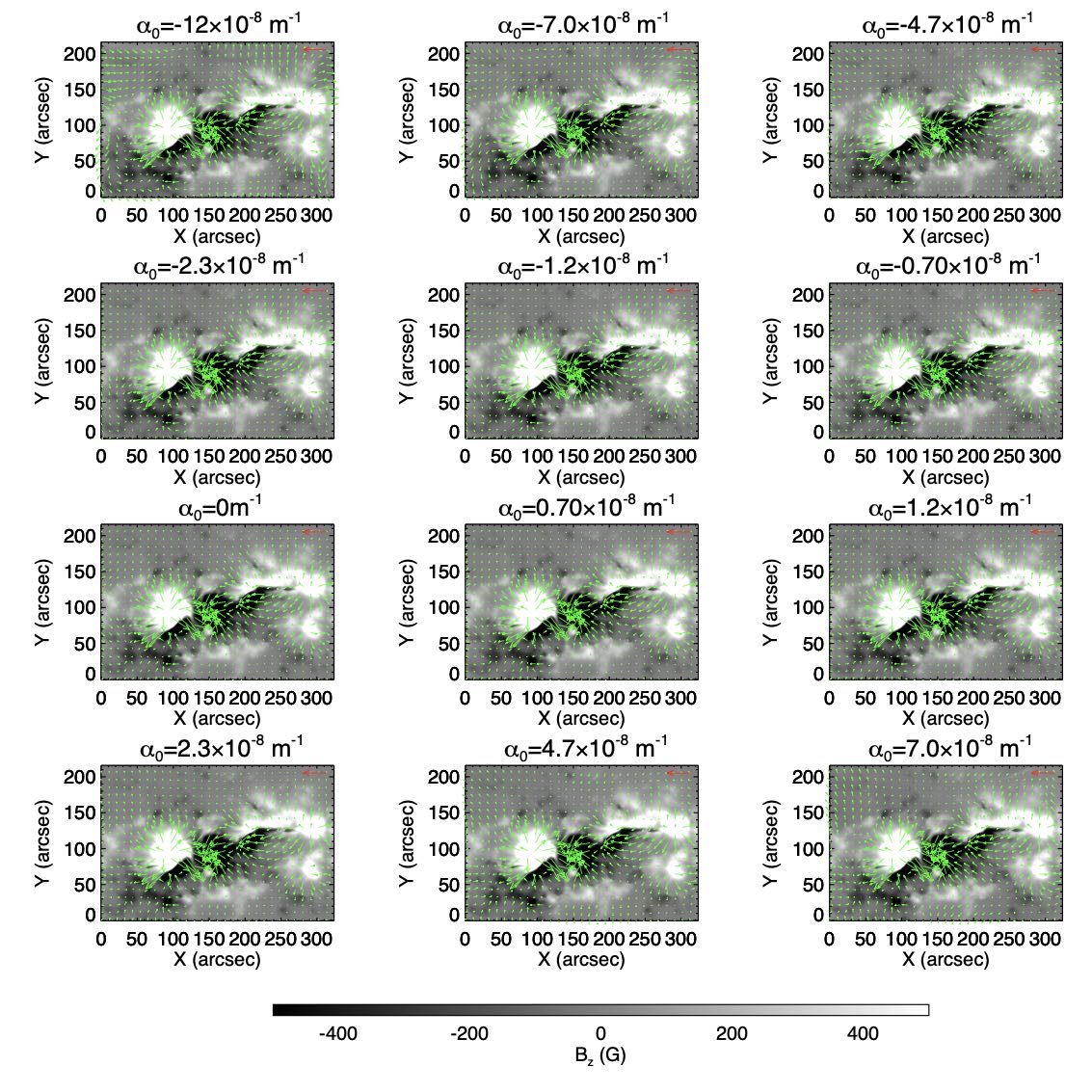}
\caption{Similar as Figure \ref{sec2:bxbybz11692_3pix}, the spatial distributions of vector magnetic field for each solution in NOAA 11967 at 2.6 Mm height. The length of the red arrow shows the field strength of 1500 G.}
\label{sec2:bxbybz11967_6pix}
\end{figure}

\begin{figure}
\includegraphics[bb=0 0 1133 1133,scale=0.45]{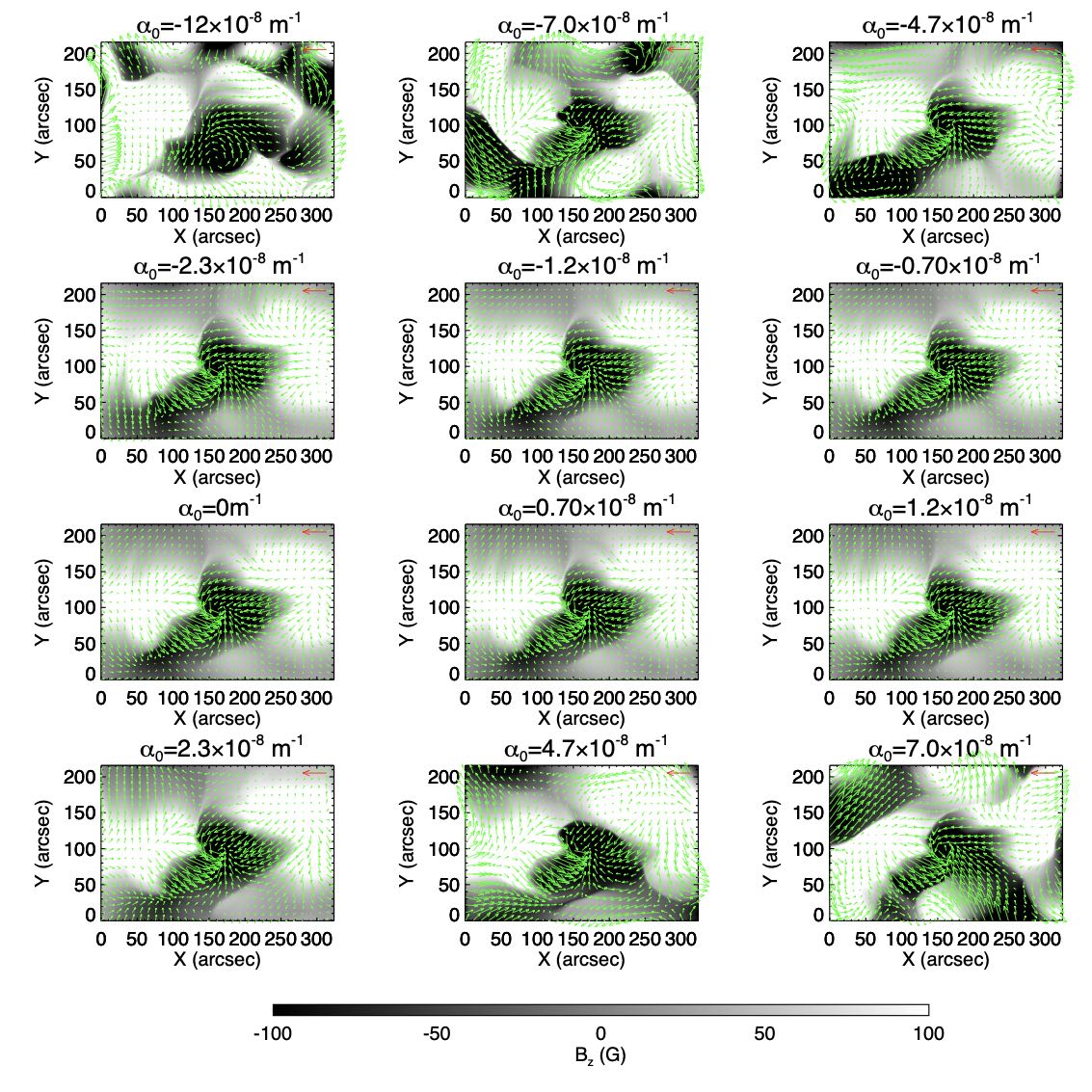}
\caption{Similar as Figure \ref{sec2:bxbybz11967_6pix}, but at 26 Mm. The length of the red arrow shows the field strength of 300 G.}
\label{sec2:bxbybz11967_60pix}
\end{figure}

\begin{figure}
\includegraphics[bb=0 0 800 579,scale=0.6]{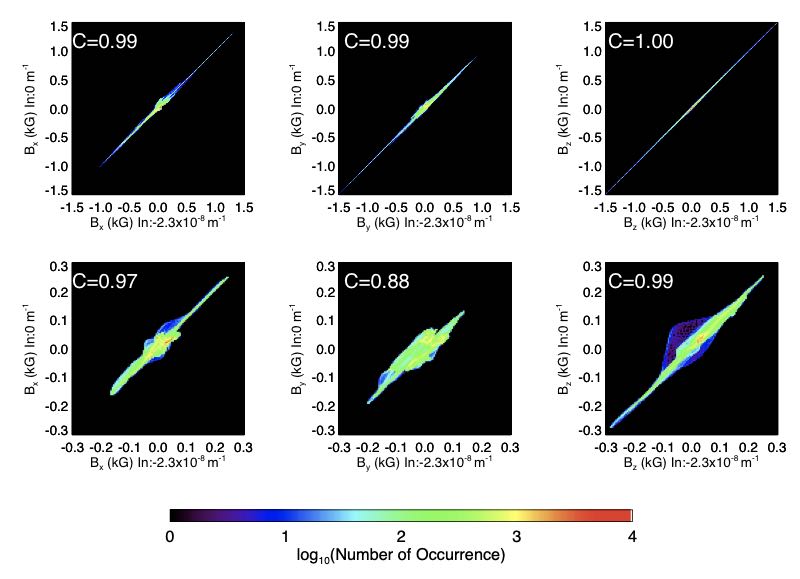}
\caption{All panels show the number density distribution in vector magnetic field from different initial conditions at 2.6 km (upper) and 26 Mm (lower) height. Comparison between $\alpha_0=0\  {\rm m^{-1}}$ and $\alpha_0=-2.3\times10^{-8}\ {\rm m^{-1}}$ in NOAA 11967.}
\label{sec2:bxbybz_2dhist_11967}
\end{figure}

\begin{figure}
\includegraphics[bb=0 0 800 579,scale=0.6]{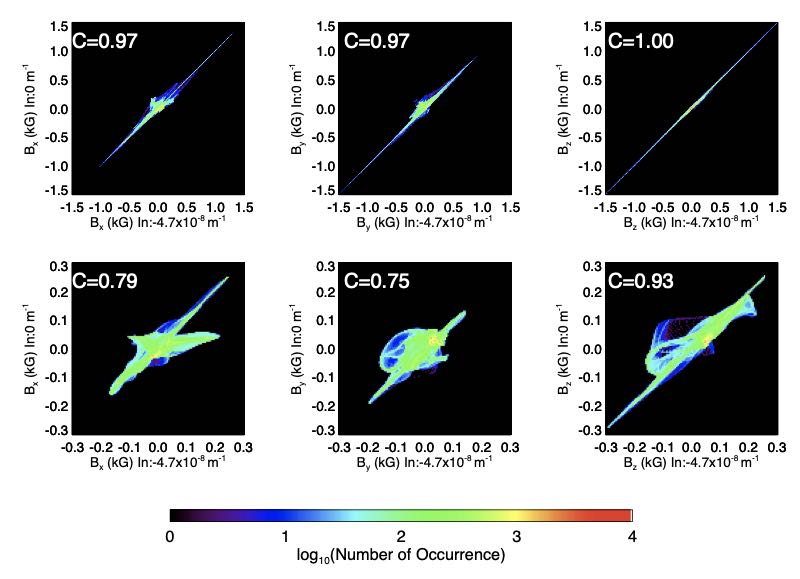}
\caption{All panels show the number density distribution in vector magnetic field from different initial conditions at 2.6 km (upper) and 26 Mm (lower) height. Comparison between $\alpha_0=0\  {\rm m^{-1}}$ and $\alpha_0=-4.7\times10^{-8}\ {\rm m^{-1}}$ in NOAA 11967.}
\label{sec2:bxbybz_2dhist_11967_m4p7}
\end{figure}

\begin{figure}
\includegraphics[bb=0 0 800 579,scale=0.6]{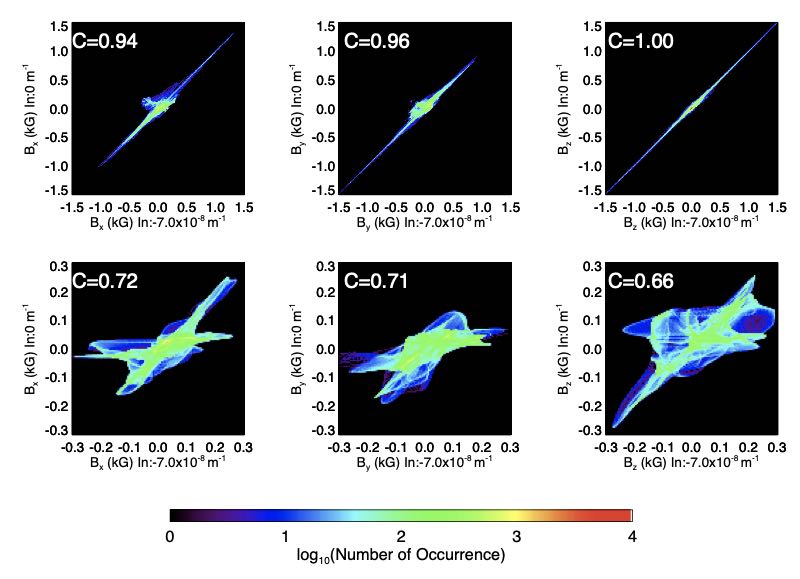}
\caption{All panels show the number density distribution in vector magnetic field from different initial conditions at 2.6 km (upper) and 26 Mm (lower) height. Comparison between $\alpha_0=0\  {\rm m^{-1}}$ and $\alpha_0=-7.0\times10^{-8}\ {\rm m^{-1}}$ in NOAA 11967.}
\label{sec2:bxbybz_2dhist_11967_m7}
\end{figure}

The spatial distributions of vector magnetic field in NOAA 11967 at 2.6 Mm and 26 Mm height are shown in Figures \ref{sec2:bxbybz11967_6pix} and \ref{sec2:bxbybz11967_60pix}.
The grayscale color shows the vertical magnetic field and green arrows show the horizontal magnetic field.
The spatial distribution of vector magnetic field is quite similar at 2.6 Mm among NLFFF results from different initial conditions.
At 26 Mm, there exist difference not only in horizontal magnetic field but also in vertical magnetic field.
Figures \ref{sec2:bxbybz_2dhist_11967},   \ref{sec2:bxbybz_2dhist_11967_m4p7},  and \ref{sec2:bxbybz_2dhist_11967_m7} show the number density distribution in vector magnetic field of NOAA 11967 with different initial condition at the height of 2.6 Mm and 26 Mm.
The result of $\alpha_0=0 \ {\rm m^{-1}}$ is compared to that with the results of $\alpha_0=-2.3\times10^{-8}\  {\rm m^{-1}}$, $\alpha_0=-4.7\times10^{-8}\  {\rm m^{-1}}$, and $\alpha_0=-7.0\times10^{-8}\  {\rm m^{-1}}$, respectively.
For all comparisons, the vector magnetic field shows strong correlation at the 2.6 Mm height, even though the initial condition is quite different.
These coefficients of NOAA 11967 are larger than those of NOAA 11692.
Similar to NOAA 11692, the small deviation can be seen in the weak magnetic field ($< 200 $ G) region.
At 26 Mm,  the number density distributions do not show strong correlation, suggesting that the NLFFF results are affected by the initial condition, although the correlation coefficients are larger than those of NOAA 11692.
The correlation coefficients become small with the comparison between the potential and  the large absolute value of initial alpha. 
However, the strong magnetic field ($>$200 G) shows good correlation even at 26 Mm height.

\subsection{Convergence in Each Calculation}
\begin{figure}
\includegraphics[bb=0 0 965 487,scale=0.5]{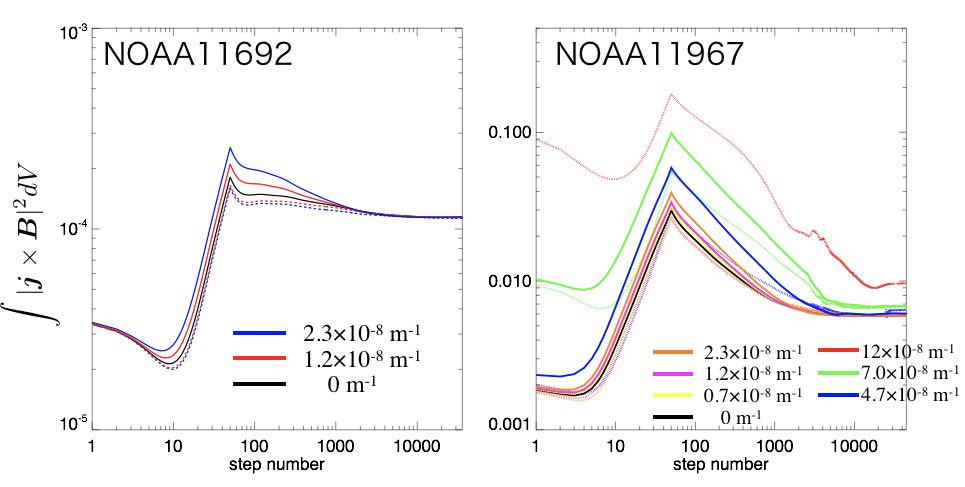}
\caption{The volume integral of the Lorenz force at each time step. Each color of the lines shows the initial absolute value of the constant force-free alpha. Solid lines correspond to the initial positive force-free alpha and dotted lines correspond to negative one.}
\label{sec2:convergence}
\end{figure}

The volume integral of the Lorentz force must vanish in the force-free condition.
However, it does not strictly vanish in numerical NLFFF computations.
Non-zero Lorentz force may be caused by the non-force-freeness at the bottom boundary and the inconsistency of the setting of the top boundary and side boundary.
For the reasons mentioned above, the Lorenz force often appears near the boundary of the calculation box.
Therefore, we usually stop the relaxation process when the volume integral of the Lorentz force converges to a certain value.
Figure \ref{sec2:convergence} shows the evolution of the volume integral of the Lorenz force as a function of time step.  
The Lorentz force is normalized by the values described in Section \ref{sec2:mhd_relaxation}. 
Each color of the lines shows the initial absolute value of the constant force-free alpha. 
Solid lines correspond to the initial negative force-free alpha, whereas dotted lines correspond to positive one. 
The Lorentz force increases between 10 and 100 step number because the electric current is transported from the bottom boundary according to Equation \ref{bottombc}.
For both active regions, the convergence speed becomes faster for negative value in comparison with the opposite value in the same absolute $\alpha_0$.
For NOAA 11692, the volume integral of the Lorentz force converges to a similar value for all initial conditions.
For NOAA 11967, when the absolute value of initial constant alpha has a smaller value, e.g., black, yellow, purple, orange,  and blue lines in Figure \ref{sec2:convergence}, the volume integral of the Lorentz force tends to smaller at the same time step compared to the large initial force-free alpha, e.g., green and red lines.  
The converged values of Lorentz force are higher than the initial values.
This is because we fixed the bottom boundary  to the non-force-free (observed) photospheric magnetic field during the late stage of the MHD relaxation. 
As a result, the Lorentz force is mainly concentrated at the lower height as shown in Appendix. 
 On the other hand, we use force-free (potnential or LFF) bottom boundary at the initial condition and the total Lorentz force is not so large in the initial stage.

\section{Discussions
\label{sec2:discussion}}

We calculated NLFFF with 5 and 12 different initial conditions for NOAA 11692 and 11967, respectively.
Summary of our results is as follows.
\begin{itemize}
\item According to the comparison with the soft X-ray image, the NLFFF shows better correspondence  in the simple active region NOAA 11692, when the initial constant force-free alpha is the same sign with the global alpha calculated from the photospheric magnetic field, as shown in Figure \ref{sec2:nlfff11692}. 
On the other hand, in the complex multi-pole active region NOAA 11967, the results of the NLFFF extrapolations are less dependent on the initial condition, as shown in Figure \ref{sec2:nlfff11967}.
\item Total magnetic energy of the NLFFF extrapolation does not strongly depend on the initial condition as shown in Tables \ref{sec2:table:metrics_11692} and \ref{sec2:table:metrics_11967}. 
The dependence of the free energy is larger compared to the total magnetic energy.
\item The solution of NLFFF at the region where the strong magnetic field exists, e.g., magnetic field in the lower height, tends to be less affected by the initial condition, as shown in Figure \ref{sec2:galpha}. On the other hand, the region in the weak magnetic field around the polarity inversion line, tends to be affected by the initial condition.
\item Except for the case $\alpha_0=-12 \times 10^{-8} {\rm m^{-1}}$ in NOAA 11967, the NLFFF extrapolation is considered to be converged, as shown in Figure \ref{sec2:convergence}. The increase of the calculation steps may not affect the results in this study.
\end{itemize}

  From Figure \ref{sec2:galpha}, the magnetic field in the lower height region tends to be less affected by the initial condition, while the magnetic field at the higher region is strongly affected by the initial condition.
There is  a possibility that this result is caused by the convergence problem of the NLFFF modeling,  however this issue is ruled out by an analysis of the Lorentz forces as a function of height, as shown in the Appendix.

 Our results suggest that NLFFF models calculated using relaxation methods, including the method used in this study, permit different results (depending on the initialization field used) in the upper regions of the computational domain. In other words, there exist several local minima for the force-free equilibrium in the higher regions. 
Because the initial condition is force-free everywhere except at the bottom boundary, a larger degree of freedom is allowed at the higher region compared to the lower region.
For example, in our MHD relaxation method, the information of the bottom boundary is transported by the propagation of the psuedo-Alfv\'en wave as mentioned above.
The region which is far from the boundary tends to be less affected by the psuedo-Alfv\'en wave and remain almost unchanged from the initial equilibrium state (potential-like or LFFF-like).
Therefore, the NLFFF extrapolation can not reproduce the magnetic field of active regions well such as have twisted structures in the corona, but have potential-like photospheric magnetic field.
Note that this possibility does not prove that the degree of freedom of the NLFFF solution is mathematically larger at the higher region because certain amount of the Lorentz force remains in our calculation.
To solve this problem, we suggest that additional observational limitation should be given to the current NLFFF modeling such that the magnetic field at the higher region can converge to the NLFFF result, which is consistent with X-ray and/or EUV imaging observations.

We compared NLFFF results with coronal loops observed with {\it Hinode}/XRT.
 For NOAA 11967, 3D magnetic field configuration matches well, and appears more consistent with X-ray observation when the absolute value of initial force-free alpha is smaller.
On the other hand, NOAA 11692 shows strong initial condition dependence in Figure \ref{sec2:nlfff11692}.
The clear difference between NOAA 11967 and 11692 is the complexity of the photospheric magnetic field.
Our results suggest that the NLFFF result of  NOAA 11967 is less affected by the initial condition than that of NOAA 11692.
Because we analyze only two active region, we can not determine the cause of this result.
Therefore, we discuss the candidates of the cause of difference in terms of  the property of two active regions.
 There are five possibilities to explain this cause.
\begin{enumerate}
\item 
NOAA 11967 has more magnetic flux, $8.4\times10^{22}$ Mx, than that of NOAA 11692, $3.7\times 10^{22}$ Mx.
This suggests that there are more strong magnetic field region above the photosphere.
\item Unbalanced magnetic flux at the photospheric height may also be the cause.
In the NLFFF modeling, magnetic flux unbalance at the bottom boundary may produce inconsistent results and may be related to the initial condition dependence.  
The net vertical magnetic flux normalized by the total magnetic flux is -0.03 for NOAA 11692 and 0.17 for NOAA 11967.
This shows that the magnitude of unbalance of vertical magnetic flux is larger in NOAA 11967 than in NOAA 11692.
Because the NLFFF results of NOAA 11967 are less affected by the initial condition, the flux unbalance in this study may not affect the initial condition dependence.
\item The force-freeness at the photospheric height is different between the active regions.
Although force-freeness of the both active regions is relatively small, that of NOAA 11967 ($|F_z|/F_p=0.089$) is smaller than that of NOAA 11692 ($|F_z|/F_p=0.17$). 
The large value of the force-freeness may affect the dependence of the initial condition on the NLFFF result.
\item The length scale of the field lines may affect the results.
While the height of the loop top in NOAA 11692 is around 100 Mm as shown in Figure \ref{sec2:nlfff11692_difangle}, that in NOAA 11967 is around 30 Mm as shown in Figure \ref{sec2:nlfff11967_difangle}.
This result means that the height of the coronal loops identified in the soft X-ray images are different between NOAA 11692 and 11967. 
Our result shows that the magnetic field in the higher region tends to be affected more strongly by the initial condition than that in the lower region.
Therefore, in comparison with the soft X-ray image in NOAA 11692, we compared the field lines, which tend to be affected by the initial condition.
However, only the difference of the length scale of the coronal loops can not explain the difference of the initial condition dependency.
As shown in Figures \ref{sec2:bxbybz_2dhist_11692} and Figures \ref{sec2:bxbybz_2dhist_11967}, NOAA 11692 is more dependent on the initial condition than NOAA 11967 even in the same height.
This result indicates that only the difference of the length scale is not the cause of the difference of the dependence between the two active regions.
\item There is a difference of the horizontal field distribution between the two active regions.
As shown in Figure \ref{sec2:fig:observation}, strong horizontal magnetic field occupies a large part of the FOV of NOAA 11967 for all four polarities.
On the other hand, in NOAA 11692, while the horizontal magnetic field can be seen in the negative sunspot, there is no strong horizontal magnetic field in the positive polarity.
As shown in Figure \ref{sec2:histo_hor_vertical}, while 6.3 $\%$ of the horizontal magnetic field in the FOV is larger than 1000 G for NOAA 11967, 0.49 $\%$ is larger than 1000 G for NOAA 11692.
This result indicates that in relaxation process, it is difficult to give sufficient electric currents (non-potential magnetic field) to the upper atmosphere in the case of NOAA 11692 because there are less currents and forces in the photosphere.
As a result, there are strong dependency on initial condition in the NLFFF modeling in NOAA 11692.
In NOAA 11967 case, the active region has strong horizontal field, which give strong perturbations to the upper atmosphere in the relaxation process.
This may be why the NLFFF modeling of NOAA 11967 converged to a consistent solution with different initial conditions.
\end{enumerate}

As shown in Tables \ref{sec2:table:metrics_11692} and \ref{sec2:table:metrics_11967},
the total magnetic energy does not strongly depend on the initial condition.
On the other hand, the free energy shows larger difference among each solution than the total magnetic energy.
Since the free energy is defined as the deviation between the magnetic energy and potential magnetic energy, the small difference of the total magnetic energy becomes large difference in free energy.
This ratio is smaller than previous studies focusing on other dependences, such as method dependence \citep[][total energy: 1.9 between ${\rm Reg^{+}}$ and ${\rm Am1^{-1}}$]{2009ApJ...696.1780D}, instrument dependence between {\it Hinode} and {\it SDO} \citep[][total energy: $\sim 2.2$, free energy: $\sim 2.5$]{2013ApJ...769...59T}, and spatial resolution dependence \citep[][total energy:$\sim 1.4$, free energy: $\sim 2$ in magnetofrictional method]{2015ApJ...811..107D}.
This means that the initial condition dependence of the total magnetic energy and free energy in our MHD relaxation method is comparatively small. 
The uniqueness of total energy and free energy can be explained by using our results.
Since the magnetic energy and free energy concentrate in the lower height, they become less dependent to the initial condition.

 Our study shows that the NLFFF results determined from relaxation methods may be strongly affected by the initial condition in the higher levels of the computational domain. We therefore recommend checking whether field lines from NLFFF models are consistent with coronal images. Our results also indicate that the sigmoidal structure can be well reproduced by changing initial condition even when the photospheric magnetic field has less twist. Regarding NOAA 11692, our NLFFF modeling produce more consistent results with X-ray observations, when we use the initial values of  $\alpha_0=-2.3\times 10^{-8} {\rm m^{-1}}$. 
This value has the same sign of the global alpha in the photosphere ( $\alpha_g= -1.0 \times 10^{-8} {\rm m^{-1}}$). 
Although the global alpha does not strictly correspond to the global magnetic twist when the magnetic field is not force-free, the sign of the global alpha gives the sign of global shear signed angle \citep{2009ApJ...702L.133T}.
Therefore, our results offer an important suggestion that the photospheric global alpha can be used for the rough estimation of the initial condition for better NLFFF modeling.

\section{Summary
\label{sec2:summary}}
Summarizing our findings, 
\begin{description}
\item[(1)] The solution of NLFFF at the region where the strong magnetic field exists, e.g., magnetic field in the lower height ($< $10 Mm), tends to be less affected by the initial condition, although the Lorentz force is concentrated at the lower height.
\item[(2)] Total magnetic energy of the NLFFF extrapolation does not strongly depend on the initial condition. 
\item[(3)] The NLFFF extrapolation of the complex active region NOAA 11967 is less dependent on the initial condition compared to that of NOAA 11692. 
\end{description}
We proposed the problem whether completely different solutions with the same bottom boundary exist or not.
We conclude that we obtain completely different 3D NLFFF structure from the different initial conditions with the same bottom boundary. 
However, the initial condition dependence is small (the correlation coefficient $C>0.9$) where the the magnetic field is strong, e.g., in the lower height ($<$10 Mm).
We also reveal that the 10-100 times larger Lorentz force, which is normalized by the square of the magnetic field strength, remains at the lower height ($<$ 10 Mm) than that at higher region ($>$ 10 Mm).
The magnitude of the dependence is also different between the two active regions.


{\bf Acknowledgements}
We appreciate the feedback offered by the anonymous referee. 
We would like to thank Tatsu-hiko Miura, Yota Watanabe, Yuichi Ike, Yohei Ema, Tomoya Kenmochi, Saiei Matsubara, Tsuyoshi Yoneda, Yusaku Tiba, and Takahito Kashiwabara for informative discussions. 
{\it Hinode} is a Japanese mission developed and launched by ISAS/JAXA, collaboratoing with NAOJ as a domestic partner, NASA and STFC (UK) as international partners. It is operated by these agencies in cooperation with ESA and NSC (Norway). We gratefully acknowledge the {\it SDO}/HMI team for providing data. Our calculations of NLFFF modeling were performed on JAXA Supercomputer System generation 2 (JSS2). This work was supported by MEXT/JSPS KAKENHI Grant Numbers JP15H05814.

\bibliographystyle{apj}


\clearpage


\appendix
\section{Analysis of Lorentz force}
\begin{figure}
\includegraphics[bb=0 0 793 904,scale=0.6]{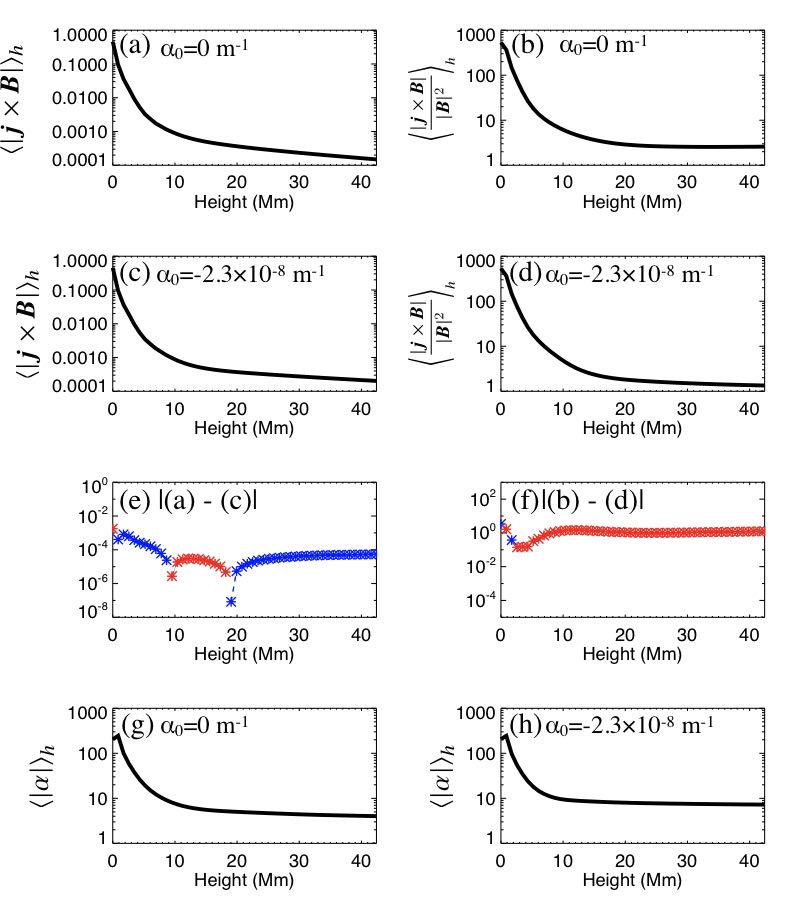}
\caption{(a) and (c): The height distribution of the Lorentz force for NOAA 11692, which is averaged at each height for $\alpha_0=0 \ {\rm m^{-1}}$  and $\alpha_0=-2.3\times10^{-8} \ {\rm m^{-1}}$, respectively. 
This result is at 24000 step. 
(b) and (d): The height distribution of the normalized Lorentz force. (e) and (f): The difference of the height distribution of the Lorentz force and the normalized Lorentz force between (a) and (c), and (b) and (d), respectively. Red asterisks show (a)$>$(c) or (b)$>$(d), while blue asterisks show (c)$>$(a) or (d)$>$(b).
(g) and (h): The height distribution of the absolute value of the force-free alpha, which is averaged at each height for $\alpha_0=0 \ {\rm m^{-1}}$  and $\alpha_0=-2.3\times10^{-8} \ {\rm m^{-1}}$, respectively. }
\label{sec2:lorentz_height_compare_noaa11692}
\end{figure}

 Although the volume integral of the Lorentz force looks converging in Figure \ref{sec2:convergence}, there is a possibility that the magnetic field has not yet converged at the higher region  because the Lorentz force may be mostly concentrated at the lower height in the calculation box.
We investigate the height distribution of the Lorentz force.
The panels (a) and (c) of Figure \ref{sec2:lorentz_height_compare_noaa11692} shows the height distribution of the Lorentz force for NOAA 11692, which is averaged at each height for $\alpha_0=0 \ {\rm m^{-1}}$ and $\alpha_0=-2.3\times10^{-8} \ {\rm m^{-1}}$, respectively.
This is the result at 24000 step.
As expected, the Lorentz force is mainly concentrated at the lower height for both solution.
The panels (b) and (d) of Figure \ref{sec2:lorentz_height_compare_noaa11692} show the Lorentz force, which is normalized by the square of the magnetic field strength and is averaged at each height for (a) $\alpha_0=0 \ {\rm m^{-1}}$  and (b) $\alpha_0=-2.3\times10^{-8} \ {\rm m^{-1}}$. 
For both $\alpha_0=0 \ {\rm m^{-1}}$ and $\alpha_0=-2.3\times10^{-8} \ {\rm m^{-1}}$, the normalized Lorentz force is also mainly concentrated at the lower height.
Below 10 Mm, the normalized Lorentz force is of the order of 10-100, while above 10 Mm, the normalized Lorentz force is of the order of 1-10.
The panels (e) and (f) show the difference of the height distribution of Lorentz force between (a) and (c),  and (b) and (d), respectively. 
Red asterisks show (a)$>$(c) or (b) $>$(d), while blue asterisks show (c)$>$(a) or (d) $>$(b).
The difference of the Lorentz force is around $10^{-4}$.
The Lorentz force of $\alpha_0=-2.3\times10^{-8} \ {\rm m^{-1}}$ is larger above 20 Mm.
At most of the height, the normalized Lorentz force of  $\alpha_0=-2.3\times10^{-8} \ {\rm m^{-1}}$ is smaller than that of $\alpha_0=0 \ {\rm m^{-1}}$.
Although the difference of the normalized Lorentz force is larger at the lower height, the ratio to the normalized Lorentz force is smaller compared to the higher region.
At the lower height, the difference of the normalized Lorentz force is of order of 0.1-1, while the normalized Lorentz force is  of order of 10-100.
The ratio of the difference is 1\% at the lower height.
On the other hand, at the higher region, the difference of the normalized Lorentz force is of order of 0.1, while the normalized Lorentz force is of order of 1.
The ratio of the difference is 10\% at the higher region.
The panels (g) and (f) show the height distribution of the absolute value of the force-free alpha, which is averaged at each height for $\alpha_0=0 \ {\rm m^{-1}}$  and $\alpha_0=-2.3\times10^{-8} \ {\rm m^{-1}}$, respectively. 
Note that the force-free alpha and the normalized Lorentz force have the same unit.
The force-free alpha shows similar distribution with the Lorentz force.
At the lower height, the force-free alpha is of order of 10-100, while at the higher region, the force-free alpha is of order 1-10.
The force-free alpha is smaller than the normalized Lorentz force at the lower height, while larger at the higher region.
This result indicates that the normalized Lorentz force is sufficiently small at the higher region, while at the lower height, there exists the significant normalized Lorentz force.

\begin{figure}
\includegraphics[bb=0 0 638 801,scale=0.7]{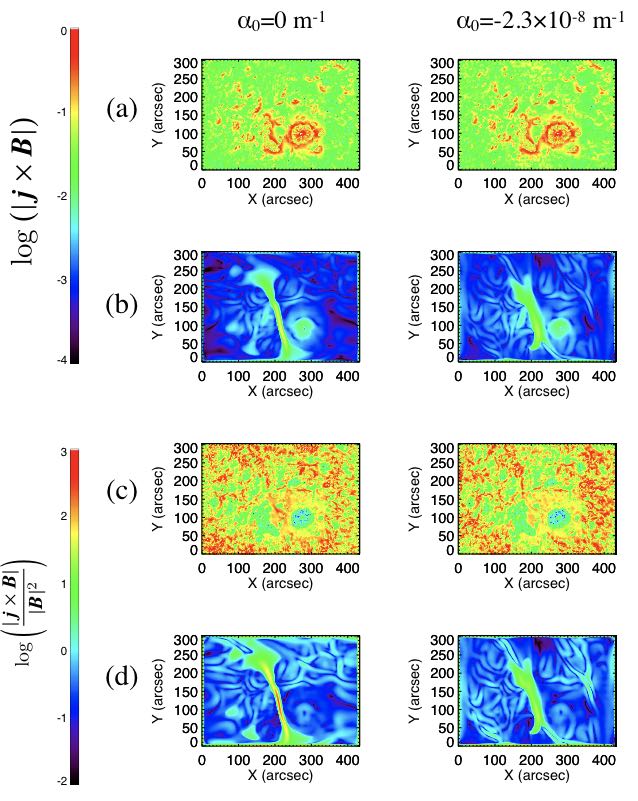}
\caption{(a) and (c): The Lorentz force and the normalized Lorentz force distribution at 2600 km height for NOAA 11692 for $\alpha_0=0 \ {\rm m^{-1}}$ (left) and $\alpha_0=-2.3\times10^{-8} \ {\rm m^{-1}}$ (right), respectively.
(b) and (d): The Lorentz force and the normalized Lorentz force distribution at 26 Mm height for NOAA 11692 for $\alpha_0=0 \ {\rm m^{-1}}$ (left) and $\alpha_0=-2.3\times10^{-8} \ {\rm m^{-1}}$ (right), respectively.}
\label{sec2:lorentz_compare_noaa11692}
\end{figure}

The panels (a) and (c) of Figure \ref{sec2:lorentz_compare_noaa11692} show the Lorentz force and the normalized Lorentz force distribution at 2600 km height for NOAA 11692 for $\alpha_0=0 \ {\rm m^{-1}}$ (left) and $\alpha_0=-2.3\times10^{-8} \ {\rm m^{-1}}$ (right). 
The panels (b) and (d) show the Lorentz force and the normalized Lorentz force distribution at 26 Mm height for NOAA 11692 for $\alpha_0=0 \ {\rm m^{-1}}$ (left) and $\alpha_0=-2.3\times10^{-8} \ {\rm m^{-1}}$ (right).
At 2600 km height, the strong Lorentz force is concentrated around the negative sunspot because the strong magnetic field is concentrated.  
The normalized Lorentz force is small in the strong magnetic field region such as negative spot umbra, while the normalized Lorentz force is large in the spot penumbra and weak magnetic field region. 
At 2600 km height, there are little difference between the distribution of  $\alpha_0=0 \ {\rm m^{-1}}$ and $\alpha_0=-2.3\times10^{-8} \ {\rm m^{-1}}$.
At 26 Mm height, the distribution of the Lorentz force and the normalized Lorentz force are similar.
This indicates that the magnetic field is more uniformly distributed at the higher region.
The Lorentz force is concentrated in the polarity inversion line.
At 26 Mm, the Lorentz force shows a slight different distribution between $\alpha_0=0 \ {\rm m^{-1}}$ and $\alpha_0=-2.3\times10^{-8} \ {\rm m^{-1}}$.
The NLFFF of $\alpha_0=0 \ {\rm m^{-1}}$ has the stronger Lorentz force in the polarity inversion line compared to that of $\alpha_0=0 \ {\rm m^{-1}}$.

\begin{figure}
\includegraphics[bb=0 0 1024 768,scale=0.5]{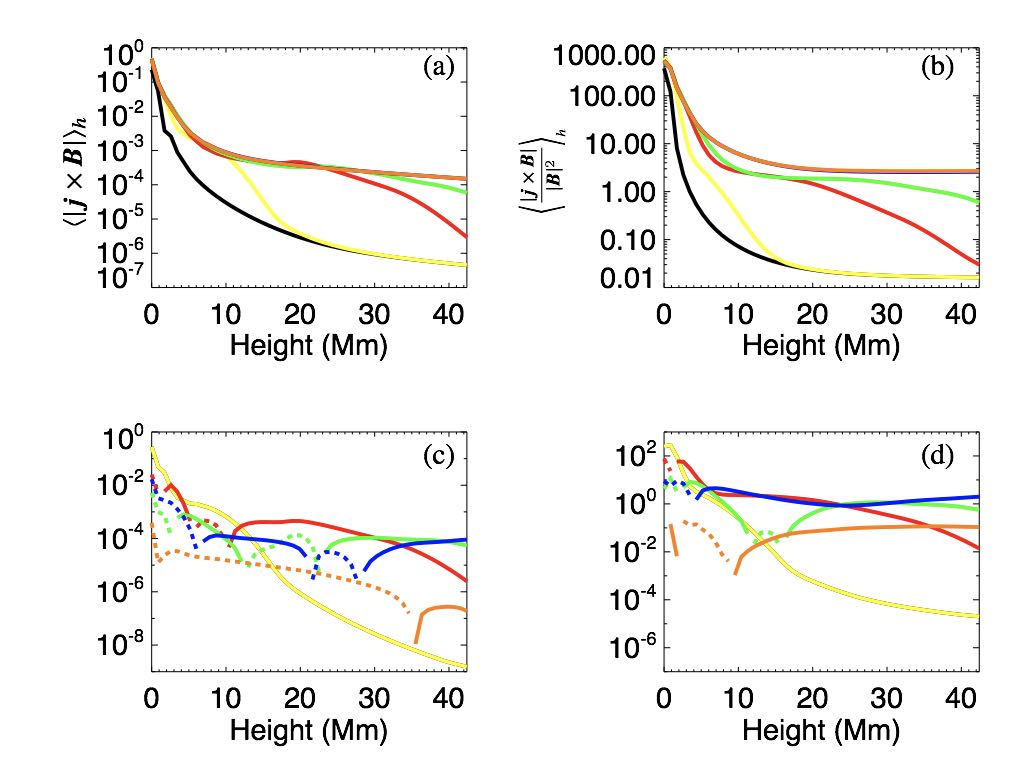}
\caption{(a) and (b):Solid lines show the height distribution of the Lorentz force  and the normalized Lorentz force for NOAA 11692, which is averaged at each height, respectively. 
This is the result of $\alpha_0=0 \ {\rm m^{-1}}$. 
Colors show each time step at 20 (black),100 (yellow), 500 (red), 900 (green), 24000 (blue), and 60000 (orange) step, respectively. 
(c) and (d): The temporal variation of the height distribution of the Lorentz force and the normalized Lorentz force. 
Colors show the time interval between 20 and 100 step (yellow), 100 and 500 step (red), 500 and 900 step (green), 900 and 24000 step (blue), and 24000 and 60000 step (orange), respectively. 
Solid lines show the increase, while the dashed lines show the decrease.}
\label{sec2:lorentz_iter_noaa11692}
\end{figure}

The panels (a) and (b) of Figure \ref{sec2:lorentz_iter_noaa11692} show the height distribution of the Lorentz force and the normalized Lorentz force for NOAA 11692, which is averaged at each height. 
These are the results of $\alpha_0=0 \ {\rm m^{-1}}$. 
Colors show each time step at 20 (black),100 (yellow), 500 (red), 900 (green), 24000 (blue), and 60000 (orange) step, respectively. 
The panels (c) and (d) show the temporal variation of the height distribution of the Lorentz force and the normalized Lorentz force. 
Colors show the time interval between 20 and 100 step (yellow), 100 and 500 step (red), 500 and 900 step (green), 900 and 24000 step (blue), and 24000 and 60000 step (orange), respectively. 
Solid lines show the increase, while the dashed lines show the decrease.
Both the Lorentz force and the normalized Lorentz force show the similar behavior.
As clearly seen, the Lorentz force and the normalized Lorentz force are transported from the lower height to the higher region with increasing the calculation step.
Between the step of 24000 (blue) and 60000 (orange), the Lorentz force and the normalized Lorentz force show little change, which means that the magnetic field does not change significantly during this steps.

\begin{figure}
\includegraphics[bb=0 0 862 768,scale=0.5]{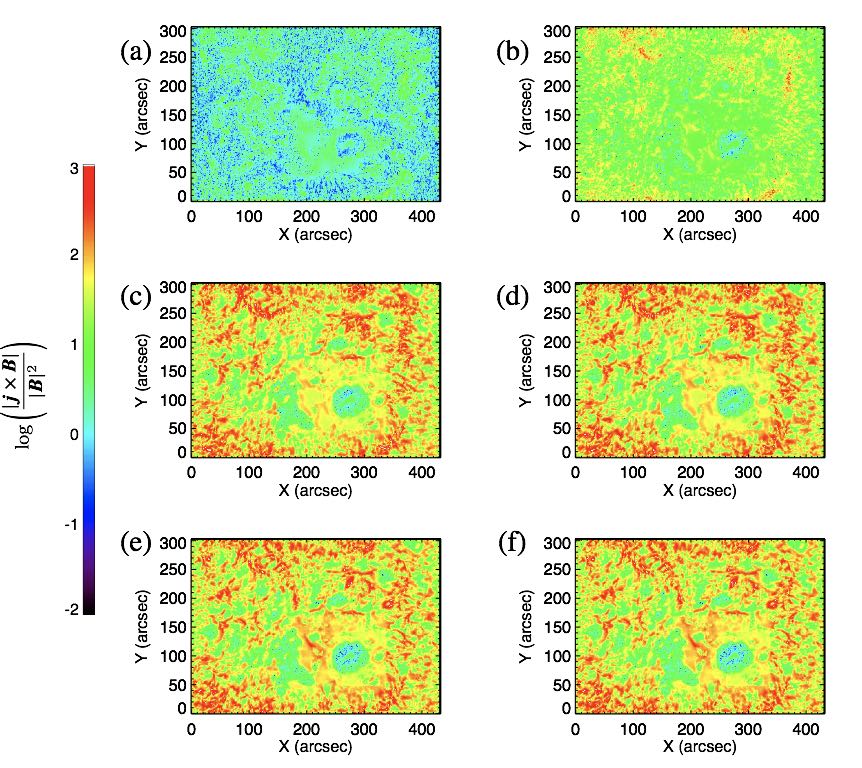}
\caption{The temporal evolution of the normalized Lorentz force distribution at 2600 km for NOAA 11692. 
This is the normalized Lorentz force of $\alpha_0=0 \ {\rm m^{-1}}$. 
Each panel shows each time step at (a) 20 step, (b)100 step, (c) 500 step, (d) 900 step, (e) 24000 step, and (f) 60000 step, respectively.}
\label{sec2:lorentz_iter_2600km_noaa11692}
\end{figure}

Figure \ref{sec2:lorentz_iter_2600km_noaa11692} shows the temporal evolution of the normalized Lorentz force distribution at 2600 km for NOAA 11692. 
This is the result of the normalized Lorentz force of $\alpha_0=0 \ {\rm m^{-1}}$.
Each panel shows each time step at (a) 20 step, (b)100 step, (c) 500 step, (d) 900 step, (e) 24000 step, and (f) 60000 step, respectively.
The normalized Lorentz force at 2600 km increases between 20 and 500 steps, while the distribution of the Lorentz force does not change significantly after 500 steps.
This result indicates that the magnetic field at the lower height does not change significantly from early calculation step, although the normalized Lorentz force remains to a certain amount.

\begin{figure}
\includegraphics[bb=0 0 862 768,scale=0.6]{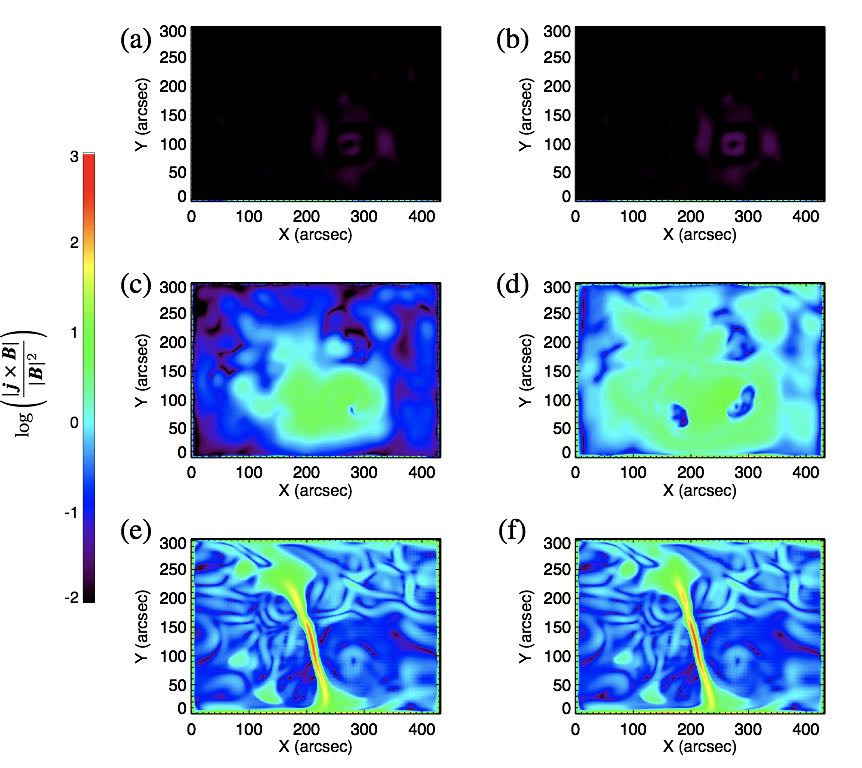}
\caption{The temporal evolution of the normalized Lorentz force distribution at 26 Mm for NOAA 11692. 
Each panel shows each time step at (a) 20 step, (b)100 step, (c) 500 step, (d) 900 step, (e) 24000 step, and (f) 60000 step, respectively.}
\label{sec2:lorentz_iter_26Mm_noaa11692}
\end{figure}

Figure \ref{sec2:lorentz_iter_26Mm_noaa11692} shows the temporal evolution of the normalized Lorentz force distribution at 26 Mm for NOAA 11692. 
Each panel shows each time step at (a) 20 step, (b)100 step, (c) 500 step, (d) 900 step, (e) 24000 step, and (f) 60000 step, respectively.
At the time step of 20 step and 100 step, there is little normalized Lorentz force because the normalized Lorentz force has not been transported from the bottom boundary as seen in Figure \ref{sec2:lorentz_iter_noaa11692}.
After 500 step, the normalized Lorentz force increases at 26 Mm and the distribution of the normalized Lorentz force does not change significantly after 24000 step.

\begin{figure}
\includegraphics[bb=0 0 792 905,scale=0.6]{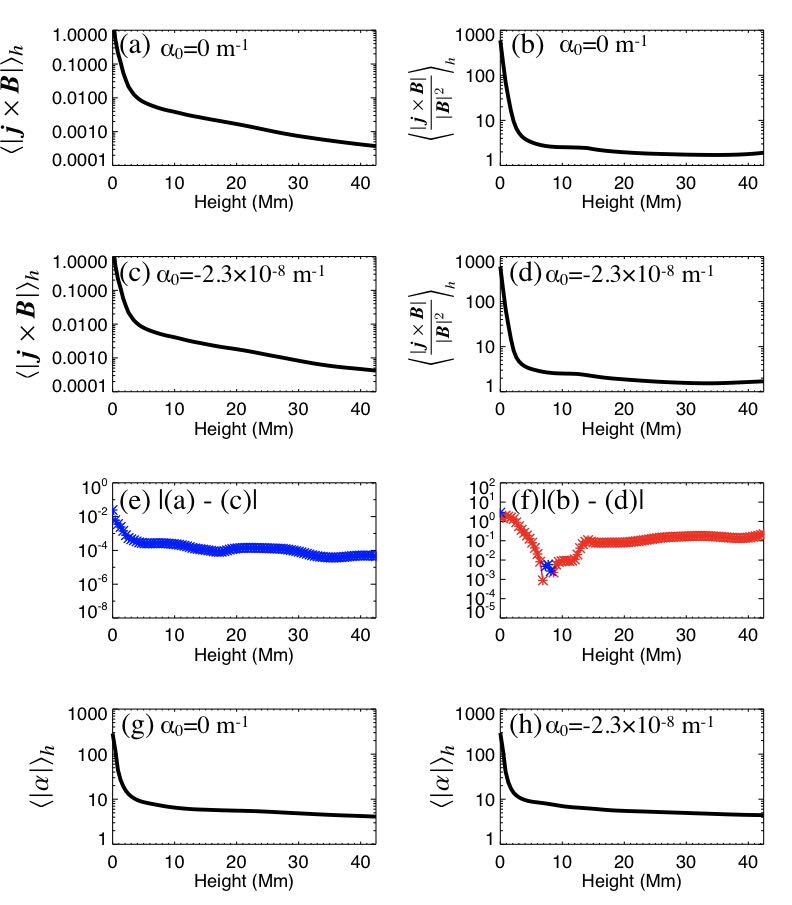}
\caption{(a) and (c): The height distribution of the Lorentz force for NOAA 11967, which is averaged at each height for $\alpha_0=0 \ {\rm m^{-1}}$ and $\alpha_0=-2.3\times10^{-8} \ {\rm m^{-1}}$. 
This is the result at 25000 step.
(b) and (d): The height distribution of the normalized Lorentz force. 
(e) and (f): The difference of the height distribution of the Lorentz force and the normalized Lorentz force between (a) and (c), and (b) and (d), respectively. 
Red asterisks show (a)$>$(c) or (b)$>$(d), while blue asterisks show (c)$>$(a) or (d)$>$(b).}
\label{sec2:lorentz_height_compare_noaa11967}
\end{figure}

\begin{figure}
\includegraphics[bb=0 0 638 801,scale=0.7]{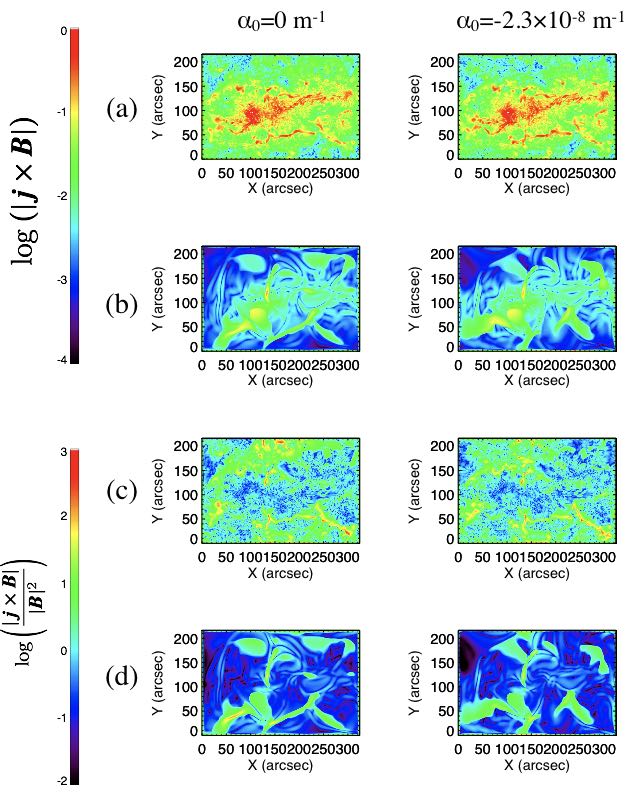}
\caption{(a) and (c): The Lorentz force and the normalized Lorentz force distribution at 2600 km height for NOAA 11967 for $\alpha_0=0 \ {\rm m^{-1}}$ (left) and $\alpha_0=-2.3\times10^{-8}\  {\rm m^{-1}}$ (right), respectively.
(b) and (d): The Lorentz force and the normalized Lorentz force distribution at 26 Mm height for NOAA 11967 for $\alpha_0=0 \ {\rm m^{-1}}$ (left) and $\alpha_0=-2.3\times10^{-8} \ {\rm m^{-1}}$ (right), respectively.}
\label{sec2:lorentz_compare_noaa11967}
\end{figure}

\begin{figure}
\includegraphics[bb=0 0 1024 768,scale=0.5]{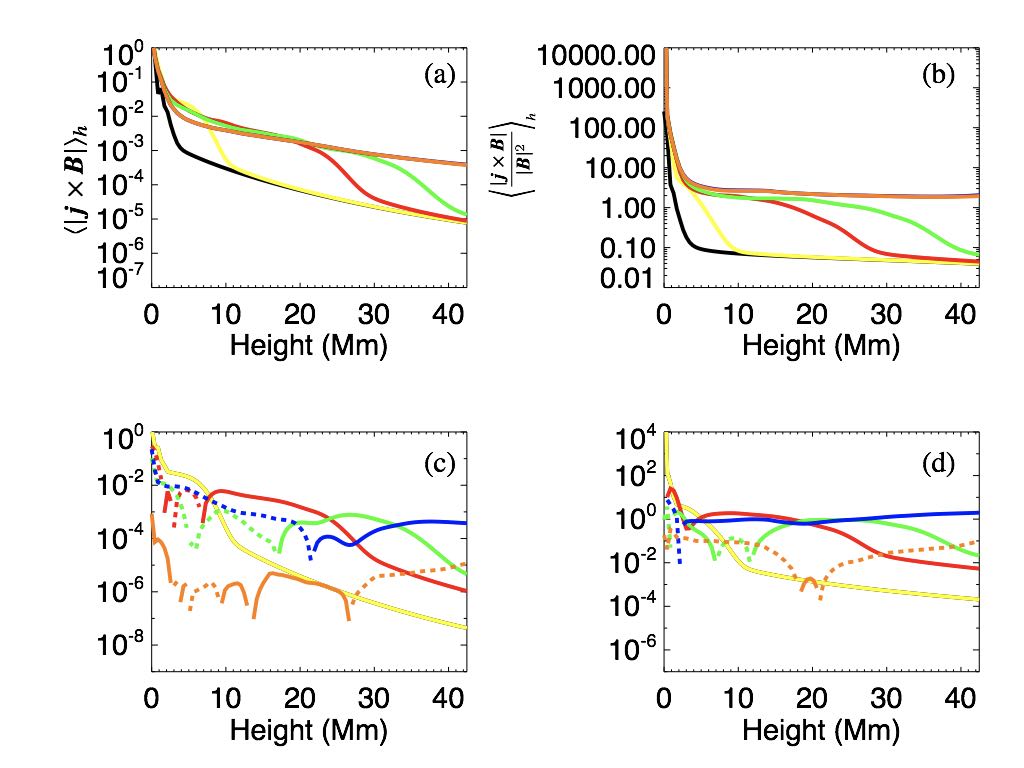}
\caption{(a) and (b):Solid lines show the height distribution of the Lorentz force and the normalized Lorentz force for NOAA 11967, which is averaged at each height. 
This is the result of $\alpha_0=0 \ {\rm m^{-1}}$.
Colors show each time step at 20 (black), 100 (yellow), 500 (red), 900 (green), 25000 (blue), and 45000 (orange) step, respectively. 
(c) and (d):  The temporal variation of the height distribution of the Lorentz force and the normalized Lorentz force. 
Colors show the time interval between 20 and 100 step (yellow), 100 and 500 step (red), 500 and 900 step (green), 900 and 25000 step (blue), and 45000 step (orange), respectively. 
Solid lines show the increase, while the dashed lines show the decrease.}
\label{sec2:lorentz_iter_noaa11967}
\end{figure}

\begin{figure}
\includegraphics[bb=0 0 862 768,scale=0.6]{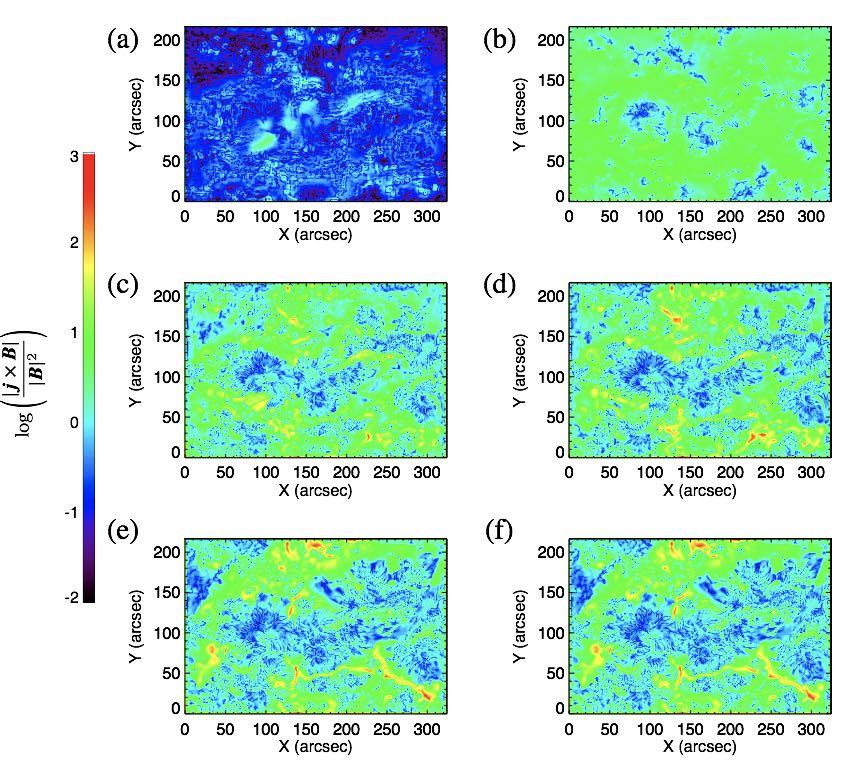}
\caption{The temporal evolution of the normalized Lorentz force distribution at 2600 km for NOAA 11967. 
This is the result of $\alpha_0=0 \ {\rm m^{-1}}$. 
Each panel shows each time step at (a) 20 step, (b)100 step, (c) 500 step, (d) 900 step, (e) 25000 step, and (f) 45000 step, respectively.}
\label{sec2:lorentz_iter_2600km_noaa11967}
\end{figure}

\begin{figure}
\includegraphics[bb=0 0 862 768,scale=0.6]{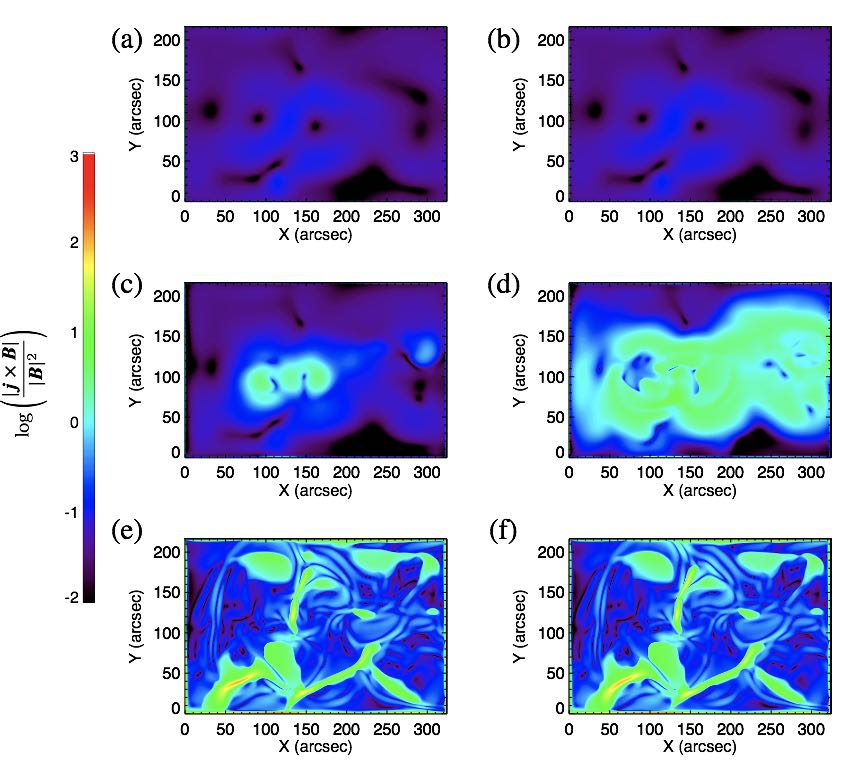}
\caption{The temporal evolution of the normalized Lorentz force distribution at 26 Mm for NOAA 11967. 
This is the result of $\alpha_0=0 \ {\rm m^{-1}}$.
Each panel shows each time step at (a) 20 step, (b)100 step, (c) 500 step, (d) 900 step, (e) 25000 step, and (f) 45000 step, respectively.}
\label{sec2:lorentz_iter_26Mm_noaa11967}
\end{figure}

Figures \ref{sec2:lorentz_height_compare_noaa11967}, \ref{sec2:lorentz_compare_noaa11967}, \ref{sec2:lorentz_iter_noaa11967}, \ref{sec2:lorentz_iter_2600km_noaa11967}, and \ref{sec2:lorentz_iter_26Mm_noaa11967} show the results of the analysis of the Lorentz force for NOAA 11967 in the same way with NOAA 11692.
The results are similar as those of NOAA 11692.

From the above results, there are two possible reasons to explain the stronger dependence of the initial condition on the NLFFF modeling at higher region.
 The first possible reason is that the information of the bottom boundary has not reached at the higher region yet.
In the MHD relaxation method, the NLFFF is achieved by the propagation of the disturbance as psuedo-Alfv\'en wave, which is produced by the artificial change of the bottom boundary according to Equation (\ref{bottombc}).
Because we assume $\rho=|\vector{B}|$ in the MHD calculation, the Alfv\'en velocity is $B/\sqrt{4\pi\rho}\sim\sqrt{B}$.
At the higher region and weak magnetic field region, there is a possibility that the psuedo-Alfv\'en wave has not reached yet.
 This possibility is rejected by the result in Figures \ref{sec2:lorentz_iter_noaa11692} and \ref{sec2:lorentz_iter_noaa11967}, which show that the Lorentz force is transported to the higher region.

The second possibility is the difference of the convergence speed among each height.
We used the resistivity defined as Equation (\ref{resistivity}), which becomes large when the Lorentz force or the velocity becomes large.
Because the Lorentz force and the velocity tends to become large at the lower height, the resistivity tends to be larger in the lower region than in the higher region.
The large resistivity allows the magnetic field to be relaxed faster at the lower height.
For the second possibility, it is unlikely that the magnetic field at the higher region converges to one unique solution by increasing iteration steps.
Regarding the convergence of the NLFFF results, from the result in Figure \ref{sec2:lorentz_iter_26Mm_noaa11692} and \ref{sec2:lorentz_iter_26Mm_noaa11967},  the distribution of the Lorentz force does not significantly change after 24000 step for NOAA 111692 and 25000 step for NOAA 11967.
This result indicates that the current scheme can not reduce the Lorentz force at the higher region any more.

We have to note that although the magnetic field at the lower height is less affected by the initial condition, certain amount of the Lorentz force remains, as shown in Figures \ref{sec2:lorentz_height_compare_noaa11692} and \ref{sec2:lorentz_height_compare_noaa11967}.
The main reason of the remaining Lorentz force in the calculation box may be due to the non-force-freeness of the photospheric magnetic field at the bottom boundary.
Since the photospheric magnetic field is not ideally force-free, the non-zero Lorentz force will be produced in the current NLFFF scheme.
Therefore, strictly speaking, we do not obtain ideal force-free solution in the current NLFFF scheme.
We will investigate how this non-force-freeness at the bottom boundary affect the accuracy of the NLFFF modeling by comparing the NLFFF and chromospheric observations (Kawabata et al., in prep).
This problem should be solved by using more force-free bottom boundary, such as chromospheric magnetic field.
\clearpage




\end{document}